\documentclass[10pt,journal,compsoc]{IEEEtran}
\ifCLASSOPTIONcompsoc
  \usepackage[nocompress]{cite}
\else
  \usepackage{cite}
\fi
\ifCLASSINFOpdf
\else
\fi
\usepackage{xspace}
\usepackage{url}
\usepackage{graphicx, transparent, color}
\usepackage{xcolor, tcolorbox}
\usepackage{hyperref}
\usepackage{tikz}
\usepackage{gensymb}
\usetikzlibrary{shadows}
\usepackage{color, colortbl}
\usepackage{filecontents}
\usepackage{stfloats}
\usepackage{capt-of}
\usepackage{soul}

\newcommand{\RA}[1]{#1}
\newcommand{\RB}[1]{#1}
\newcommand{\RC}[1]{#1}
\newcommand{\RS}[1]{#1}

\newcommand{\RAL}[2]{\hyperref[#1]{\RA{#2}}}
\newcommand{\RBL}[2]{\hyperref[#1]{\RB{#2}}}
\newcommand{\RCL}[2]{\hyperref[#1]{\RC{#2}}}
\newcommand{\RSL}[2]{\hyperref[#1]{\RS{#2}}}

\newcommand{\RAtext}[2]{\label{#1}\RA{#2}}
\newcommand{\RBtext}[2]{\label{#1}\RB{#2}}
\newcommand{\RCtext}[2]{\label{#1}\RC{#2}}

\newcommand{\RAback}[0]{}
\newcommand{\RBback}[0]{}
\newcommand{\RCback}[0]{}

\begin{document}

\twocolumn
\pagenumbering{arabic}
\setcounter{page}{1}
\setcounter{figure}{0}
%
\title{{INSPECT}: Intrinsic and Systematic Probing Evaluation for Code Transformers}


%
%
%
%

\author{Anjan~Karmakar
        and~Romain~Robbes%
\IEEEcompsocitemizethanks{\IEEEcompsocthanksitem Anjan Karmakar is with the Free University of Bozen-Bolzano, Italy. 
\IEEEcompsocthanksitem Romain Robbes is with the University of Bordeaux, CNRS, Bordeaux INP, LaBRI, UMR5800, Talence, France.}
\thanks{Manuscript received ----/revised ----}}

%
%

\markboth{Journal of \LaTeX\ Class Files,~Vol.~14, No.~8, August~2015}%
{Shell \MakeLowercase{\textit{et al.}}: Bare Demo of IEEEtran.cls for Computer Society Journals}
%



\IEEEtitleabstractindextext{%
\begin{abstract}
Pre-trained models of source code have recently been successfully applied to a wide variety of Software Engineering tasks; they have also seen some practical adoption in practice, e.g. for code completion. Yet, we still know very little about \emph{what} these pre-trained models learn about source code. In this article, we use \emph{probing}---simple diagnostic tasks that do not further train the models---to discover to what extent pre-trained models learn about specific aspects of source code. We use an extensible framework to define 15 probing tasks that exercise surface, syntactic, structural and semantic characteristics of source code. We probe 8 pre-trained source code models, as well as a natural language model (\texttt{BERT}) as our baseline. We find that models that incorporate some structural information (such as \texttt{GraphCodeBERT}) have a better representation of source code characteristics. Surprisingly, we find that for some probing tasks, \texttt{BERT} is competitive with the source code models, indicating that there are ample opportunities to improve source-code specific pre-training on the respective code characteristics. We encourage other researchers to evaluate their models with our probing task suite, so that they may peer into the hidden layers of the models and identify what intrinsic code characteristics are encoded.
\end{abstract}

}

\maketitle

\IEEEdisplaynontitleabstractindextext

%
\IEEEpeerreviewmaketitle

\newcommand\topspace{\rule{0pt}{3ex}}     
\newcommand\bottomspace{\rule[-1ex]{0pt}{0pt}}  
\newcommand{\myparagraph}[1]{\noindent \topspace \textbf{#1} }

\newcommand\tstrut{\rule{0pt}{1ex}}     
\newcommand\bstrut{\rule[1ex]{0pt}{0pt}}  


\newcommand{\task}[1]{\texttt{#1}\xspace}

\newcommand{\TAN}[0]{\task{TAN}}
\newcommand{\TANexp}[0]{Top AST Node}
\newcommand{\AST}[0]{\task{AST}}
\newcommand{\ASTexp}[0]{AST Node Tagging}
\newcommand{\IDT}[0]{\task{IDT}}
\newcommand{\IDTexp}[0]{Identifier Tagging}

\newcommand{\IDN}[0]{\task{IDN}}
\newcommand{\IDNexp}[0]{Identifier Tagging}

\newcommand{\KTX}[0]{\task{KTX}}
\newcommand{\KTXexp}[0]{Keyword Tagging}


\newcommand{\LEN}[0]{\task{LEN}}
\newcommand{\LENexp}[0]{Code Length}

\newcommand{\OCT}[0]{\task{OCT}}
\newcommand{\OCTexp}[0]{Operator Count Total}

\newcommand{\OCU}[0]{\task{OCU}}
\newcommand{\OCUexp}[0]{Operator Count Unique}

\newcommand{\VCT}[0]{\task{VCT}}
\newcommand{\VCTexp}[0]{Variable Count Total}
\newcommand{\VCU}[0]{\task{VCU}}
\newcommand{\VCUexp}[0]{Variable Count Unique}

\newcommand{\CSC}[0]{\task{CSC}}
\newcommand{\CSCexp}[0]{Code Structure Count}

\newcommand{\NML}[0]{\task{NML}}
\newcommand{\NMLexp}[0]{Number of Loops}

\newcommand{\NMS}[0]{\task{NMS}}
\newcommand{\NMSexp}[0]{Number of If Structures}

\newcommand{\CPX}[0]{\task{CPX}}
\newcommand{\CPXexp}[0]{Cyclomatic Complexity}
\newcommand{\NPT}[0]{\task{NPT}}
\newcommand{\NPTexp}[0]{Npath Complexity}
\newcommand{\MXN}[0]{\task{MXN}}
\newcommand{\MXNexp}[0]{Maximum Indentation}

\newcommand{\tokbased}[0]{\emph{Token-based}\xspace}
\newcommand{\countbased}[0]{\emph{Count-based}\xspace}
\newcommand{\compbased}[0]{\emph{Complexity-based}\xspace}
\newcommand{\bugbased}[0]{\emph{Mistyped}\xspace}
\newcommand{\switchbased}[0]{\emph{Replacement}\xspace}


\newcommand{\JBL}[0]{\task{JBL}}
\newcommand{\JMB}[0]{\task{JMB}}
\newcommand{\JFT}[0]{\task{JFT}}
\newcommand{\JBLexp}[0]{Jumbled Code Tokens}
\newcommand{\REA}[0]{\task{REA}}
\newcommand{\REAexp}[0]{Relational operator to Assignment}
\newcommand{\SCK}[0]{\task{SCK}}
\newcommand{\SCKexp}[0]{Switch Compatible Keyword}
\newcommand{\SRK}[0]{\task{SRK}}
\newcommand{\SRKexp}[0]{Switch Random Keyword}
\newcommand{\SRI}[0]{\task{SRI}}
\newcommand{\SRIexp}[0]{Switch Random Identifier}
\newcommand{\TYP}[0]{\task{TYP}}
\newcommand{\TYPexp}[0]{Detect Invalid Types}

\newcommand{\taskdef}[1]{
\vskip 2mm \noindent \textbf{\textit{\csname #1exp\endcsname}} (\csname #1\endcsname)
}

\newcommand{\doublecheck}[1]{
\textcolor{red}{\textbf{#1}}
}

\newcommand{\INSPECT}[0]{\texttt{INSPECT}\xspace}
\newcommand{\JEMMA}[0]{\texttt{JEMMA}\xspace}

\newcommand{\BERT}[0]{\texttt{BERT}\xspace}
\newcommand{\CodeBERT}[0]{\texttt{CodeBERT}\xspace}
\newcommand{\CodeBERTa}[0]{\texttt{CodeBERTa}\xspace}
\newcommand{\GraphCodeBERT}[0]{\texttt{GraphCodeBERT}\xspace}
\newcommand{\CodeT}[0]{\texttt{CodeT5}\xspace}

\newcommand{\PLBART}[0]{\texttt{PLBART}\xspace}
\newcommand{\CodeReviewer}[0]{\texttt{CodeReviewer}\xspace}
\newcommand{\UnixCoder}[0]{\texttt{UniXCoder}\xspace}
\newcommand{\JavaBERT}[0]{\texttt{JavaBERT}\xspace}

\newcommand{\etal}[0]{\emph{et al.}\xspace}
\newcolumntype{?}{!{\vrule width 1.2pt}}

\definecolor{res-gray}{cmyk}{0, 0, 0, 0.07, 1.00}
\newcommand{\resultbox}[1]{
	\begin{tcolorbox}[colback=res-gray,colframe=gray!30!white,top=3pt,                                                     left=3pt,right=3pt, bottom=3pt] #1
	\end{tcolorbox}
}

\newcommand{\RQ}[1]{\texttt{RQ#1}\xspace}

\NewDocumentCommand\gold{}{
    \includegraphics[scale=0.03]{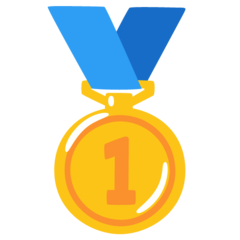}
}

\NewDocumentCommand\silver{}{
    \includegraphics[scale=0.03]{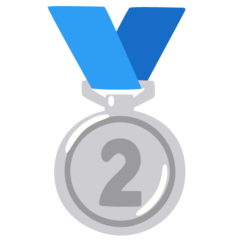}
}

\NewDocumentCommand\bronze{}{
    \includegraphics[scale=0.03]{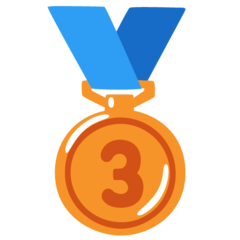}
}
\IEEEraisesectionheading{\section{Introduction}\label{sec:introduction}}

\IEEEPARstart{T}{he} outstanding success of transformer-based \cite{vaswani2017attention} pre-trained models in NLP such as \texttt{BERT} \cite{devlin2019bert}, has inspired the creation of a number of similar pre-trained models for source code. These pre-trained models are first trained on a large corpus of code in a self-supervised manner and then fine-tuned on downstream tasks. These models include sequence-based models such as \texttt{CodeBERT} \cite{feng2020codebert}, \texttt{CuBERT} \cite{DBLP:journals/corr/abs-2001-00059}, or \texttt{PLBART} \cite{DBLP:journals/corr/abs-2103-06333}, to name a few. Other models include more structural information on source code, such as \texttt{GraphCodeBERT} \cite{guo2021graphcodebert}, or \texttt{UniXCoder}\cite{guo2022unixcoder}. Finally, several recent models leverage scaling laws \cite{kaplan2020scaling} to improve performance just by virtue of their size, such as \texttt{Codex} \cite{chen2021evaluating}, \texttt{CodeGen} \cite{nijkamp2022conversational}, \texttt{InCoder}\cite{fried2022incoder} and \texttt{AlphaCode} \cite{li2022competition}. This effort is ongoing, with novel pre-trained models of source code being announced regularly.   

These large-scale pre-trained transformer models for code have been shown to perform spectacularly well on a wide range of software engineering tasks, which led to the release of a number of new tools. These include source code completion tools, such as Tabnine, IntelliCode \cite{svyatkovskiy2020intellicode}, or GitHub CoPilot, that leverages OpenAI's Codex \cite{chen2021evaluating} to propose multi-line code completion from natural language prompts. More generally, the increasing availability of these AI-based development tools has triggered discussions of their promises and concerns \cite{ernst2022ai}.

The progress made with pre-trained source code models is genuinely encouraging. However, its remains unclear \emph{what exactly do these models learn about source code}. Do they learn the tasks at hand, or, like the mathematic prodigy Clever Hans (a horse\footnote{\textit{\url{https://en.wikipedia.org/wiki/Clever_Hans}}}) \cite{johnson1911clever}, do they exploit superficial cues \cite{kavumba2019choosing} in their input?  Are source code models ``stochastic parrots'' \cite{bender2021dangers}, that overly rely on memorization \cite{karmakar2022codex}? 
There are many such questions about source code models that can be asked, all of which impact our understanding of source code models and their capabilities.

In this paper, we focus on a more specific question, namely: to what extent do pre-trained source code models learn about the specific characteristics of source code? This question is important since pre-trained models come from the field of \texttt{NLP}. However, source code and natural language are very different. For instance, source code is highly structured, and can be unambiguously parsed. Functions or methods form complex structures with an intricate control flow over many statements. Source code identifiers can be formed of multiple natural language words (e.g., \texttt{ArrayIndexOutOfBoundsException}). Source code utilizes punctuation in a way that is vastly different than natural language. Small changes to source code can lead to widely different behaviour, crashes, or compilation errors. 

In spite of this, several architectures have been applied as-is to source code. For instance, Codex is a variant of the \texttt{GPT-3} natural language model, that is further fine-tuned on source code; it inherits its tokenization, only adding a few additional tokens to better handle source code indentation. Whether the default \texttt{NLP} training objectives are as effective for structured source code as it is for less structured \texttt{NLP} applications is an open question.  On the other hand, some models, such as \GraphCodeBERT \cite{guo2021graphcodebert}, do adopt additional pre-training objectives to better account for source code's characteristics. Whether the additional training objectives are effective in learning more source code characteristics is also an open question.

Before addressing these questions however, there is first a methodological issue: \emph{how} can we evaluate specific characteristics of source code, of which there can be many? An emerging field of research addresses these with \emph{probes}. Probes are diagnostic tasks, which are designed to evaluate a specific characteristic of the input data. To probe a pre-trained model of interest, one trains a simple classifier to predict specific properties of its input (as specified by the diagnostic task), based on the \emph{frozen} vector embeddings of the pre-trained model. The degree of success in the probing tasks indicates whether the information probed for is present in the pre-trained embeddings, and to which extent. Probing has been extensively used for natural language models, and has already begun to pick up steam with numerous probing tasks \cite{alain2018understanding, shi-etal-2016-string, adi2017finegrained, belinkov-etal-2017-neural,  DBLP:journals/corr/abs-1808-08949, DBLP:journals/corr/abs-1905-06316, conneau-etal-2018-cram, tenney2019learn} investigating a diverse array of natural language properties. We provide more background about probing in the next section. 

In this work, we adapt the probing paradigm to evaluate the source code knowledge within pre-trained source code models. In Section \ref{sec:tasks} we define a set of 15 probing tasks, that cover a variety of source code characteristics. 
The tasks cover multiple characteristics of the \textit{Java} programming language, such as reasoning on identifiers, source code structure, and code correctness. We also present \INSPECT, the probing framework that we developed for this study, in Section~\ref{sec:probes:framework}; \INSPECT allows us to easily extend our probing test suite by defining new tasks, as well as to very easily probe models that are made available on the huggingface model hub. The framework is freely accessible on github at \url{https://github.com/giganticode/inspect}. We encourage other researchers to use \INSPECT as an \textbf{intrinsic benchmark} of their source code models. 

We use our tasks to probe 8 pre-trained code models, and also a natural-language model which serves as our baseline. We provide a detailed methodology of our study, and describe the models and the datasets that we use in section \ref{sec:methodology}. The models range from ones that \emph{should have no knowledge} of source code (the \texttt{BERT} model, trained on a natural language corpus), to models that should have moderate knowledge of source code (e.g., \texttt{CodeBERT} which follows \texttt{BERT}'s training procedure, but on a source code corpus), up to models that should have a more advanced knowledge of source code (e.g., \texttt{GraphCodeBERT}, which has a code-specific pre-training objective, along with data-flow information).

We present the results of our study in section \ref{sec:results}. We answer research questions related to the {general performance of the models on the probing tasks, as well as comparing the performance of different models, the performance on different categories of tasks, as well as the performance across layers}. Overall, we find while the pre-trained models encode some source code characteristics well, there is room for progress: for some tasks, even the most advanced model fail to significantly outperform our baseline\BERT, that should have no explicit knowledge of code. We also observe that models that introduce more structural source code information in their training tend to perform better, which provides a way forward. 

Finally, we discuss the results  \ref{sec:discussion} and the limitations \ref{sec:limitations} of this work in section, and conclude in section \ref{sec:conclusion}. In particular, we conclude that defining new pre-training procedure that better emphasize source code characteristics should be developed and more systematically investigated, such as via our probing framework, \INSPECT.
 
This work is an extension of a short paper previously published in ASE 2021 \cite{karmakar2021probes}. Compared to our previous work, we perform a much more extensive study. During the course of this work, we designed 19 new tasks, eventually keeping 13 in the final version, and replacing 2 of the original 4 tasks. Overall, 85\% of the tasks are new to this paper. The tasks are divided in 5 different task categories.  Having multiple tasks for each category (rather than a single one) allows us to find trends about which source code characteristics tend to be better encoded in the models (e.g., semantic characteristics tend to be better encoded than structural characteristics).  The original work evaluated 3 models and a baseline, while this version evaluates 8 models (and the baseline). Having a wider selection of models allow us to find or dismiss certain factors that might impact the quality of the encoding of the characteristics in the models (e.g., models with code-specific training objectives appear to better encode source code characteristics, whereas models that use more generic training objectives are less successful). To account for the growth in the amount of data to present (close to 95\% of the data is new to this paper), we conducted an entirely new and much more extensive analysis of the results. This analysis looks at additional research questions and sub-questions in ways that are very different from the original paper.  Following the results, we added an extensive discussion of the results, their implications, and a discussion of their relationship to the emerging literature on probing and analyzing source code models. We also expanded the background and limitations section to make the paper much more self-contained.


\section{Probing: Background and Related Work}
\label{sec:background}
We provide an introduction to the notion of probing as performed in this study. We invite readers wishing to further explore the topic, to consult the recent survey on probing \cite{belinkov2022probing}, which also discusses some advanced approaches.

\subsection{Probing Neural Networks}
The goal of probing is to assess to what extent a Deep Neural Network learns an implicit representation of specific characteristics of its training data during training. Such characteristics are varied and very domain-specific, such as:
\begin{itemize}
    \item In Neural Machine Translation (\texttt{NMT}), whether a model has some representation of the active/passive voice or tense of the original input \cite{shi-etal-2016-string}; 
    \item In Natural Language Processing (\texttt{NLP}), whether a natural language model encodes some information about the word order of a sentence \cite{adi2017finegrained};
    \item Whether a neural chess engine encodes representations of chess concepts, such as the playing side being in check, or the strength of chess pieces \cite{mcgrath2021acquisition};
    \item Whether a source code model encodes representations of a code snippet's inherent complexity \cite{karmakar2021probes}
\end{itemize}

\noindent
Note that these characteristics are usually \emph{implicitly} present in the training data, rather than being explicitly fed to the models while training. E.g., the training data for an NMT model is simply the text to translate, and does not explicitly indicates the tense or the active/passive voice of a sentence. 

\myparagraph{Representation learning.} Deep Learning models {learn} {vector representations} of their input during training. The learned representations are based on multi-dimensional internal weights, which are optimized during the learning process. Once a model is trained, the learned vector representations can be obtained by processing any input through the model and collecting the weights from the hidden state.

\myparagraph{Probing.} We use \emph{probes} to ascertain if, and to what extent, learned vector representations of pre-trained models encode specific characteristics of interest; and whether they have been implicitly learned by the models. A \emph{probe} consists of two components: 1) a \emph{probing} \textit{task} and 2) a \emph{probing classifier}. 

\myparagraph{Probing Task.} A probing task is an auxiliary diagnostic task, generally designed to analyze implicit learning in models. A probing task may require models to predict certain characteristics from the input features given that the model is not directly trained to predict them in the pre-training stage. Each probing task should be a convincing proxy for a given characteristic of interest. For instance, to verify whether a pre-trained code model has some implicit knowledge of code size, a diagnostic task could be constructed to probe on the model's (frozen) pre-trained weights whether it can classify code snippets by size. If the pre-trained vectors of the model encode such information, the probe results will reflect the same. Similarly, a diagnostic probing task could be used to probe for the concept of word order, to randomly alter half of a set of sentences by swapping two words, and classify inputs as original or altered \cite{conneau-etal-2018-cram}. 

\myparagraph{Probing Classifiers.} One critical aspect regarding probes is that \emph{the pre-trained models are not further trained on the probing tasks}. Rather, the model's learned pre-trained weights are first extracted and used to compute the vector representations of each input sample (i.e. the pre-trained model is \emph{frozen}), which are then passed on as input features to a \emph{simple} probing classifier, typically a simple linear classifier. Note that the probe does not have access to the original input, just the resulting vector representations. This makes probing for even very simple concepts still informative, as they are not directly accessible to the model.

Only the linear classifier is optimized when training on the diagnostic task. During training, the probe should learn which of its input features---if any---are relevant for the task at hand. Thus, if one or more features encode the concept of interest, the probe should be able to perform the task successfully. On the other hand, if the original model has not learnt the concept of interest during its training, the probe should fail at the task, with a performance on a test set that differs little from random chance accuracy. 

\myparagraph{Probe Complexity.}  \label{R3.C7}\RCtext{R3.C13}{One of the assumptions behind probing is that a representation of a source code characteristic should be \emph{simple}. Complex representations are more likely to contain spurious relationships (e.g., a model might notice a relationship between the length of a piece of code and its complexity, rather than modelling complexity directly). One way to enforce this is by using simple classifiers:
if a simple classifier can successfully solve the task, it means that the representation of the concept is accessible and easy to extract from the model. We use a linear layer as the probing classifier. These are similar in complexity to a standard classification head that might be attached to a pre-trained model. Note that a linear layer can only learn linear combinations of its input: this limitation imposes a constraint on the simplicity (and thus, in terms of probing, its quality) of the representation of the existing concept in the pre-trained model. Other works use a Multi-Layer Perceptron (\texttt{MLP}) with one hidden layer as a probe. These can learn more complex non-linear combinations of features. If only a complex probe can solve the diagnostic task, there is a risk that the probe itself learns to solve the task as the training progresses, rather than relying on the actual pre-trained vectors of the pre-trained models} \RCback \cite{DBLP:journals/corr/abs-1909-03368}.

\myparagraph{Summary and Source Code Example.} The goal of a probe is to test whether, and to what extent, a specific characteristic of the training data is encoded in a model's internal representation. For instance, we might want to test whether a pre-trained source code model such as \CodeBERT encodes any notion of the complexity of source code. To do so:

\begin{enumerate}
    \item We first collect a set of training/test inputs (e.g. \textit{Java} code snippets at the method-level) and labels (e.g. each method's Cyclomatic Complexity). 
    \item We then proceed to extract learned vector representations for each input sample in the task dataset by processing them through a pre-trained model. 
    \item Finally, we train a linear classifier on the extracted vector representations, and evaluate the probe on the test set. This gives us the performance metrics for each layer of the model---with predictions based principally on the learned vector representations.  
\end{enumerate}

\begin{figure}
  \includegraphics[width=0.99\linewidth]{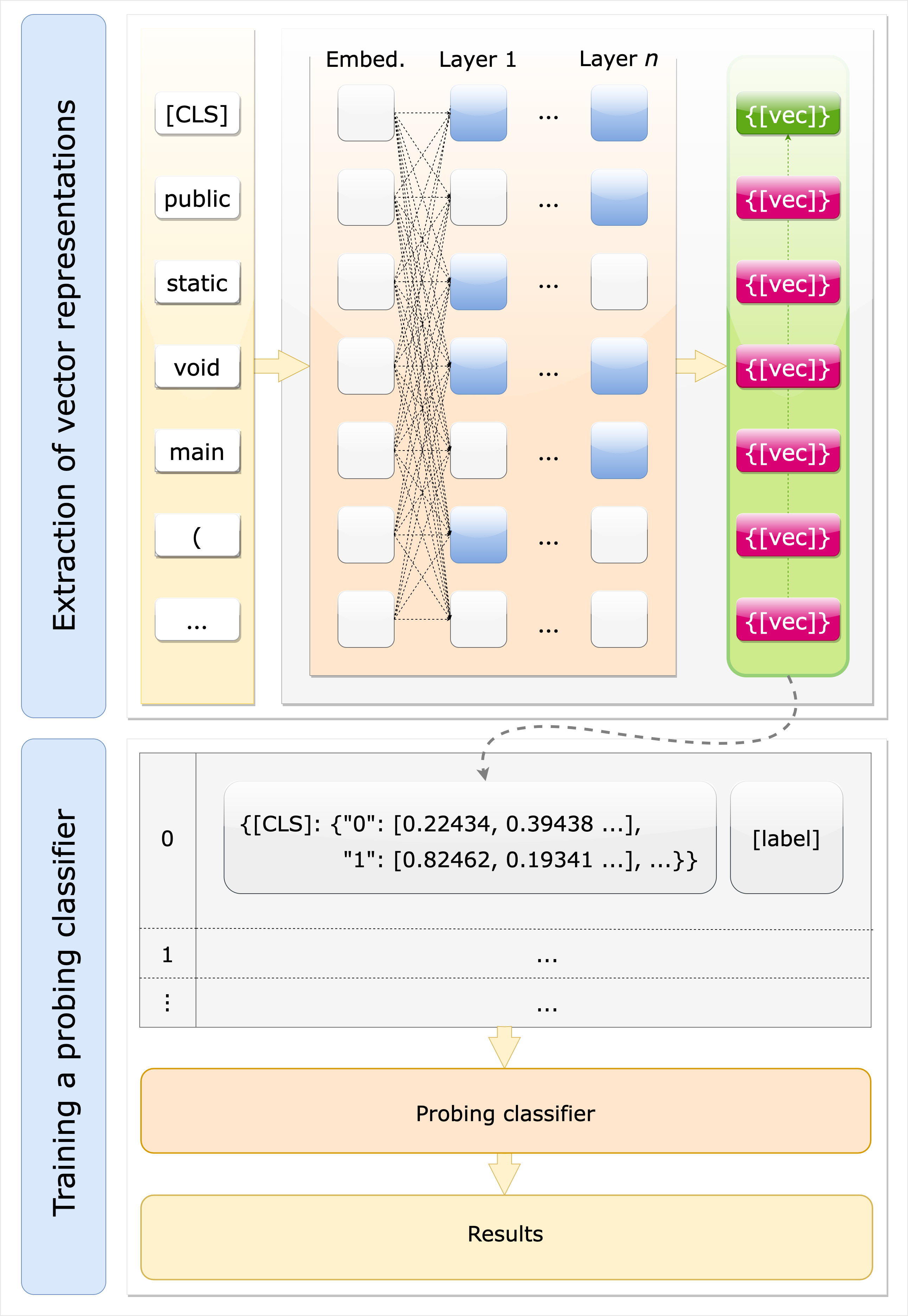}
  \caption{Overview of the probing procedure}
  \label{fig:probing_overview}
\end{figure}

Fig. \ref{fig:probing_overview} illustrates the above steps where pre-trained vector representations and corresponding property labels are used to probe for certain intrinsic information which are expected to be encoded in the model layers.

\subsection{Related Work}

\myparagraph{Probing in NLP.} An early study by Shi \etal found that \texttt{LSTM} machine translation models capture several concepts of syntax, such as voice, tense, or part of speech \cite{shi-etal-2016-string}. Belinkov \etal focused on finer-scale concepts, such as word morphology, finding that character-based machine translation models better capture morphology \cite{belinkov-etal-2017-neural}. Conneau \etal defined 10 probing tasks, including sentence length, tense, \texttt{AST} tree depth, and others, finding that the performance on probing tasks outperform several simple baselines \cite{conneau-etal-2018-cram}. The study by Jawahar \etal \cite{jawahar-etal-2019-bert} show that \BERT encodes phrase-level information in the lower layers, and a hierarchy of linguistic information in the intermediate layers, with surface features at the bottom, syntactic features in the middle and semantic features at the top of a vector space. Hewitt and Manning designed a \textit{structural probe} \cite{hewitt2019structural} that evaluates whether pre-trained models such as \BERT \cite{devlin2019bert} or \texttt{ELMo} \cite{matthew2018peters} have representations of a sentence's \emph{entire} syntax tree embedded as implicit node distances in the vector representations: they find that this is the case for medium to late layers.
Clark \etal probed \BERT's attention heads, finding that some individual heads attended to specific linguistic concepts \cite{clark2019does}.
The previous studies are but few examples: studies of the \BERT models alone have spawned a subfield known as \textit{\texttt{BERT}-ology} with over 150 studies surveyed \cite{rogers2020primer}. 

\myparagraph{Probing in other fields.} Probing has been applied beyond textual models only. DeepMind's AlphaZero game playing models have been probed for the games of chess  \cite{mcgrath2021acquisition} and Hex \cite{forde2022concepts}. For chess, the model uses a Convolutional Neural Network to encode the board; the CNN is probed for close to 150 individual chess concepts (e.g., king safety, threats to pieces). Cao \etal applied probing to Vision and Language models, that reason on images and text at the same time \cite{cao2020behind}. Recently, studies relate the hidden state of language models with fMRI data obtained from humans reading the same text the language models are processing \cite{schrimpf2021neural, caucheteux2021long, caucheteux2022deep}, finding correlations between some of the language model's activation weights, and brain activity.

\myparagraph{Probing in Software Engineering.} Our initial work \cite{karmakar2021probes} was the first dedicated study of probing for source code models. The only earlier example we are aware of uses a single coarse task (programming language identification)---and is not the focus of the paper \cite{feng2020codebert}. Since our initial work, several additional studies have emerged. 

Wan \etal analyzed two pre-trained code transformers, \texttt{CodeBERT} and \texttt{GraphCodeBERT} \cite{wan2022they}. They first analyzed the model's attention heads, and whether the attention aligns with the syntax structure of code, finding that it does. They then probed the word embeddings to find out if they embed values similar to inter-token \texttt{AST} distances, finding that it does, but that it varies among layers. Finally they checked whether they could reconstruct the \texttt{AST} from the learned representations, finding that they can to an extent.  

Other works focused on attention more specifically. Chen \etal introduced \texttt{CAT}-probing \cite{chen2022cat}. They define a metric, the \texttt{CAT}-Score, to relate the attention matrix (how the Transformer's attention for a given token attends to other tokens) for a given layer, with the distance matrix of the \texttt{AST} nodes. They find that \texttt{CAT}-scores for source code models are correlated with their performance on code summarization, and that the \texttt{CAT}-scores vary per layer and per language, with a tendency for higher scores in the earlier layers. Zhang \etal \cite{zhang2022does} define a similar metric over the attention matrix, an aggregated attention score, with which they can derive a graph of relationships between tokens.

Hernández Lopez \etal use the structural probe of Hewitt \cite{hewitt2019structural} to study various pre-trained models of source code \cite{lopez2022ast}. They find that \CodeBERT and \GraphCodeBERT better capture the \texttt{AST} than other code models such as \CodeT and \CodeBERTa, and that the \texttt{AST} representation is more encoded in the middle layers.

These works employ for the most part a holistic approach to probing source code models, primarily focusing on the entire \texttt{AST}. In contrast, we define multiple probing tasks to probe for the presence of more fine-grained concepts than the entire \texttt{AST}. \RBtext{R2.C6.1}{The closest work to ours is the work by Troshin and Chirkova }\cite{troshin2022probing} \RB{which builds up on our initial work by defining new probing tasks. They define 8 tasks which do not overlap with our 15, as we have an increased focus on metrics and incorrect code detection. In addition, space allows us to conduct a significantly more in-depth analysis of our results (which also span a broader selection of models). Another important difference is the original data used. For 6 tasks, Troshin and Chirkova use an existing dataset of 10,000 java methods with which they derive individual task datasets. The relatively small size of the data prevents the selection of subsets of the data to calibrate the difficulty of the task. For instance, the baseline for their “is variable declared” task (a constant prediction that the variable is declared) has a 90\% accuracy. The initial datasets for their two other tasks have 934 and 200 datapoints. In contrast, we start with a much larger initial dataset (with 8 million methods), allowing us to select specific subsets of the data (e.g., all our datasets are balanced). As a consequence of our different task selection and dataset construction, we tend to find lower performance for the source code models. In section} ~\ref{sec:otherstudies}, \RB{we compare our results with this study.} \RBback

Finally, Paltenghi and Pradel \cite{paltenghi2021thinking}, and Paltenghi \etal relate neural attention weights with human eye-tracking data \cite{paltenghi2022extracting}. They find some correlation between the neural attention weights of language models and eye-tracking data.


\section{The \INSPECT Task Suite}
\label{sec:tasks}

In this section we introduce our suite of probing tasks. Our probing tasks consider a variety of source code characteristics, grouped in several categories. Selecting the tasks for the probing suite among a pool of candidate tasks proved to be challenging: there are many characteristics of source code that we could select, ranging from the most basic to the most intricate. We thus selected tasks to cover a broad set of source code characteristics, while minimizing (as much as possible) the number of tasks in our suite. At the highest level, we divide the tasks into 3 broad categories:
\begin{enumerate}
    \item \textbf{Token-based tasks}, which include code token tagging tasks and total token count task (Section \ref{sec:tasks:token}).
    \item \textbf{Metrics-based tasks}, which include tasks related to identifying code structures, and unique occurrences of operators and identifiers; as well as implicit understanding of code complexity (Section \ref{sec:tasks:metrics}). 
    \item \textbf{Incorrect-code tasks}, which include switched code, misspelled, and jumbled code tasks (Section \ref{sec:tasks:incorrect}). 
\end{enumerate}

\noindent
The above tasks probe for one or more of the code characteristics with regard to syntactic, semantic, structural, or surface information. In addition to the above tasks, we discuss some candidate tasks that were considered but have not been selected for this study (Appendix \ref{appendix:tasks:candidate}). Should the need arise, our suite of probing tasks is extensible: we have developed a framework 
(Section \ref{sec:probes:framework}) 
to this effect, to help users conduct extended probing studies on further models.

\subsection{Token-based tasks}
\label{sec:tasks:token}
Tokens are the most basic code elements a pre-trained model can reason on. For instance, in order for a pre-trained model to be good at code completion, it must necessarily learn and interpret the syntactic formation of a sequence of code tokens, and predict a syntactically valid next token. Thus, the capacity of source code models to infer the type of individual tokens is an elementary source code characteristic that we want to specifically evaluate. We first define two tasks that specifically target individual tokens: one for the specific case of identifiers (\IDN), and another for the other types of tokens (\KTX). Subsequently, we define a task to probe for the size of the input, e.g. number of tokens (\LEN).  

\taskdef{IDN} Source code identifiers make up most of source code, up to 70\% according to some estimates \cite{deissenboeck2006concise}. Further, there are coding conventions that influence the format of the identifiers according to its role (e.g. a class name is different from a variable or method name). \IDN determines whether pre-trained models have sufficient syntactic and semantic knowledge to distinguish among the different types of user-defined identifiers. Pre-trained models must classify input source code tokens as one of four types of identifiers (package, class, method, and variable names). \RBtext{R2.C3}{Unlike most other tasks, we feed a single identifier to the model, rather than a method; the probing classifier uses the \texttt{[CLS]} token as usual. We do this because it is uncommon to find package names in method source code; instead, we collect package names from import declarations, which are much more common}. \RBback

\taskdef{KTX} This task evaluates the capacity of pre-trained models to discriminate between various types of keywords, operators, and symbols. The goal of the task is to learn the differences between types of keywords (e.g., \textit{float} is in a different category than \textit{return}).  We define several classes for keywords, operators, and symbols. We divide keywords into 4 keyword classes, e.g. modifiers (\texttt{private}, \texttt{protected}, \texttt{public}), flow control (e.g., \texttt{for}, \texttt{if}, \textit{return}), etc. We divide operators in 5 classes e.g. arithmetic, assignment, relational, etc. We also add \textit{Java} symbols, in a single class \RCtext{C3.14}{(e.g., {``\{"}, {``;"}, {``]"})}. \RCback 

We divide each of the classes in training, validation, and test sets to specifically probe for the ability of the models to generalize. For instance, for primitive types, we might have: \textit{int}, \textit{double}, \textit{char}, \textit{boolean} in the training set; \textit{string}, \textit{long} in the validation set, and \textit{float}, \textit{short} in the test set. Thus the probe should look for features that describe the \emph{categories} of tokens in the learned representation, rather than specific tokens. For this task we extract the representation of the specific keyword, symbol, or operator token in question, corresponding to the label, rather than the aggregated input representation, i.e. the \texttt{[CLS]} token.

\taskdef{LEN} Length is the most basic attribute we can probe for which may be implicitly learned by the models given any input. It is also the least dependent on source code knowledge, relying on the number of tokens, and as thus is useful to frame the performance of the other tasks. We measure code length in tokens. Since method sizes vary widely, we define 5 classes by binning the values into 5 different labels. Note that predicting the length of a code snippet may appear trivial, but this is not the case: since we probe the vector representation of a single summary token from the models (see section \ref{sec:methodology}), the amount of information the models can use may be limited.

\subsection{Metrics-based tasks}
\label{sec:tasks:metrics}
Since many models are trained and evaluated on tasks that concern methods (e.g., method naming, method summarization), the method is the next most natural unit to examine. This category of tasks is based on the ability of models to detect quantifiable characteristics of source code. 

Since metrics are numeric values, we could have formulated theses tasks as regression tasks. We ultimately chose to formulate them as classification tasks for two reasons: in order to be more consistent with the other tasks of the suite--comparing models across all tasks in terms of accuracy, and, to reduce the impact of confounding factors. For example, since some metrics can be correlated to code size, a model that predicts, e.g, a high complexity for all larger methods, could score deceptively well in a regression scenario. \RCtext{R3.C2}{We observed this during the design of this study when comparing regression and classification variants of the same task}\RCback. By defining classes and requiring an exact prediction on those classes, we reduce the incentive of models to rely on factors that might be indirectly correlated.

We subdivide the metrics-based task into two sub-categories: a) count-based metrics, which probe for surface-level metrics of the method, that provide general indicators on the code understanding of the model (\OCU, \VCU, \CSC), and b) complexity-based metrics, which measure various aspects of a method's complexity (\MXN, \CPX, \NPT). 

The count-based metric tasks describe basic characteristics of source code. There could be several alternative metrics that we could use in this category. We focused on metrics describing the most general concepts, and ultimately settled on operators (\textit{+}, \textit{*}, etc.) (\OCU), and variables (\VCU), and code structures (\textit{if}, \textit{for}, \textit{switch} etc.) (\CSC). 

With the complexity-based metric tasks, we analyze more advanced structure-based and control-flow based reasoning in pre-trained models. Source code complexity involves having a detailed view of the entire method. Since there are various measures of code complexity, we probe for three different aspects of varying difficulty. From easiest to hardest, our tasks are: \MXN, \CPX, and \NPT.

\taskdef{OCU} 
Operators are one of the most elemental way to manipulate data in source code. This task probes the models for their representations of the concept of source code operators. We count the number of \emph{unique} operators present in the method (e.g., multiple assignment operators will be counted only once). Thus, the task emphasizes more operator diversity than simply counting operators, and is less affected by code size. We define 10 classes representing the exact count of unique operators, \RBtext{R2.C4}{without binning}\RBback.

\taskdef{VCU} 
Identifiers are omnipresent in source code. \VCU probes the models for their representations of identifiers at the level of a method. Similarly to \OCU, we count the number of \emph{unique} identifiers in the method, rather than the total count, to emphasize diversity rather than counting. Likewise, we define 10 classes that represent the exact count of unique variables, without binning. 

\taskdef{CSC} This task probes for a more comprehensive view of control flow in the method. Here, we count all the code structures present in the method (e.g., if, for, do-while, while, switch or try blocks). Thus this task goes beyond \NMS and \NML as it probes for additional structures. For \CSC we define 10 classes, representing the total count of code structures in the method, without binning.

\taskdef{MXN} We first test for a \emph{proxy} of source complexity, which is the depth of indentation. Indentation often reflects complexity as code blocks inside other code blocks are, by convention, further indented to the right. Indeed, Hindle \etal found correlations between indentation level and various complexity metrics \cite{hindle2008reading}.  We define 5 classes, as deeper indentation levels are uncommon. 

\taskdef{CPX} Cyclomatic complexity is an established measure of complexity: it counts the number of linearly independent paths in the control flow graph \cite{mccabe1976complexity}. This requires a more advanced level of reasoning as it requires not only knowing about the existence of code structures, but also how they relate to one another and how they might affect control flow. We measure Cyclomatic Complexity to define 10 classes corresponding to exact complexity values ranging from 0 to 9, without binning \RBback.

\taskdef{NPT} Npath complexity measures the number of possible execution paths in a method, excluding cycles \cite{nejmeh1988npath}. Compared to Cyclomatic complexity, this metric has a much higher spread of values as the number of possible paths grows faster than the number of linearly independent paths. For a pre-trained model, we expect that reasoning about Npath complexity is thus more complex than Cyclomatic complexity. \RBtext{R2.C5.2}{To accommodate for the larger range of values and NPT's exponential behavior, we use binning to manually define 10 classes. The bins progressively widen: 1, 2, 3, 4--6, 7--8, 9--10, 11--15, 16--20, 21--30, 31--100.
}\RBback

\subsection{Incorrect-code tasks} 
\label{sec:tasks:incorrect}

Models trained on existing source code should have some notion of what correct code looks like. Indeed there is evidence that buggy source code is less \textit{\textbf{natural}} than fixed source code for n-gram models \cite{ray2016naturalness} and for \texttt{LSTM}s \cite{karampatsis2020big}. For the final category of tasks, we probe the ability of model to differentiate between code that is correct and incorrect in various ways. Most of the incorrect code samples in these tasks would be easily caught at compile time. Thus, we use these probes to determine whether pre-trained models understand such inconsistencies in source code inputs. 

We divide the incorrect-code tasks into two sub-categories: a) mistyped code probes (\TYP, \REA, \JBL), and b) semantic replacement probes (\SRI, \SRK, \SCK), which provide general indicators on the degree of syntactic and semantic understanding of the models.

\taskdef{TYP} For this task, negative samples have a single primitive datatype that is intentionally misspelled (e.g., \textit{float} to \textit{flaot}, or \textit{folat}). This allows us to probe to which extent models are sensitive to mistypes that would be easily detected by a compiler. For each negative sample, we misspell a single datatype, while there may be other correct occurrences of the same datatype in the sample. 

\taskdef{REA}
The Relational to Assignment Operator mistyped task evaluates whether pre-trained models can distinguish between the syntactic usages of relational and assignment operators. Some code snippets are mutated so that the relational operators are switched with assignment operators (e.g. \texttt{<=} may be switched with \texttt{+=}, or \texttt{!=} with \texttt{/=}). The mutated code may look superficially correct, therefore, such a change may be difficult to detect.

\taskdef{JBL}  For this task, the invalid code snippets are mutated by swapping two consecutive source code tokens, chosen randomly (e.g., \texttt{int foo = 4;} to \texttt{int = foo 4;}). This task probes the sensitivity of models to the order of tokens: this is a small change at the level of a method, but a model that has knowledge of source code syntax should perform well in this task.

\taskdef{SRI} This is a semantic token replacement task where one occurrence of a valid identifier is replaced with {another} non-identical identifier from the same sample chosen randomly. The mutated code looks superficially correct: without additional information (e.g., which class, variables, and methods are in scope). Thus \SRI probes models for their ability to determine if an identifier is \textit{out of context}.

\taskdef{SRK} Invalid code snippets are mutated by replacing a language keyword such as \textit{int}, with any random keyword such as \textit{for}. Thus the probability that the token is not compatible to the context is high. This probes the models for their sensitivity to language rules---syntax and semantics---in a specific context.

\taskdef{SCK} Invalid code snippets are mutated by replacing a language keyword such as \textit{double}, with another compatible keyword from the same keyword category such as \textit{int} (primitive datatypes). Another example is switching \textit{assert} with \textit{throws}, both of which belong to the same keyword category (error handling).


\section{The \INSPECT Probing framework}
\label{sec:probes:framework}

\INSPECT is the framework that we make available for researchers to perform probing studies. \INSPECT is written in Python but it is programming language agnostic---it can work with datasets with samples in multiple programming languages. In our case, we use the \JEMMA Dataset \cite{karmakar2022jemma} to define probing tasks for the \textit{Java} programming language.

\myparagraph{Task definition.} We use the \JEMMA Dataset, derived from \texttt{50K-C} Dataset \cite{8595165_50KC}, in conjunction with the \JEMMA Workbench to gather data from a large corpus of \textit{Java} source code. Once a dataset is created, \INSPECT automates the rest of the work related to probing evaluation. We have a set of 15 probing tasks that can be readily used to probe further models. However, since tasks can vary widely depending on the code characteristic of interest and programming language of interest, users may also define additional tasks. 

\myparagraph{Get frozen representations.} The first step that \INSPECT automates is getting the learned representations from the pre-trained models. \INSPECT makes it easy to interface with models that are on the HuggingFace model hub, the most common way to share Transformer-based models\footnote{\textit{\url{https://huggingface.co/models}}}. Given a pre-trained model url, \INSPECT will download it, pass each task sample (from training, validation, and test sets) through the model, and collect the model's activations at each layer. 

\myparagraph{Train Probes.} \INSPECT then uses the vector frozen representations as input for each probe sample and collects the corresponding labels from the original datasets. \RCtext{R3.C4}{Following which, \texttt{INSPECT} trains and evaluates the probing classifier on the probing tasks. The original model is not fine-tuned during this step}\RCback

\myparagraph{Compute results.} \INSPECT collects performance data of the probe from each model layer. The framework generates extensive data visualizations of the model's performance, including raw and derived performance metrics exported as \texttt{CSV}s for further analysis, heatmaps of the model's per-layer performance, and confusion matrices of predictions. 

\myparagraph{Time and space constraints.} Once a new probing task is defined, the process is entirely automated. Depending on the task, dataset size, and number the models to probe on, the amount of space (to store the frozen representations), and time (to train the probe) needed may vary. In our case, the amount of space needed to store the frozen representations for each task is about 9-12 GB, considering just the representations from the models that are included in our evaluation. We estimate that running a single probe (in our case, a probing task with a sample size of 10k) on 8 source code models requires roughly 6 hours, including pre-trained feature extraction, \RC{hyperparameter tuning}, and probing classification, on a single NVIDIA 2080Ti GPU. The time and space considerations multiply, when more probing tasks (and/or models) are considered. 

\myparagraph{Availability} \INSPECT is open-source and is available on GitHub at: \textit{\url{https://github.com/giganticode/inspect}}. All probing task datasets used in this paper are also available in the same repository. We encourage researchers to evaluate their models with our probing suite, and to extend our suite with new probing task datasets exploring further characteristics.


\section{Methodology}
\label{sec:methodology}

\subsection{Datasets, metrics and preprocessing}
\myparagraph{Datasets.} To build the datasets for our probing tasks, we retrieved our code samples parsed at the method-level from the \texttt{JEMMA} Dataset. 
For each task dataset we carefully selected 10,000 suitable \textit{Java} method samples, taking care that all of the classes are balanced (1,000 samples per class if the task has 10 classes; 2,000 if it has 5; 5,000 for binary classification). \RCtext{R3.C6}{We exclude basic Java getters and setters to make the task less reliants on these easy methods. When possible, we select methods that are shorter, to avoid truncation; however some rare labels do not make this always possible. For models that have a window of 512 tokens, on average, 3\% of methods are truncated}. We also define a smaller dataset for each task, with 1,000 samples only, to study the impact of less data. For each probing task we split the data into training, validation, and test datasets in the following ratio: 60-20-20.

\myparagraph{Performance metrics.} Since all of our tasks are balanced, we use classification accuracy as the standard metric in this study. Due to the large number of tasks, accuracy also allows us to simplify the presentation compared to using multiple metrics such as recall, \texttt{MSE}, or other regression metrics.

We considered using regression for the metrics task but after judicious assessment elected not to do so, both for consistency and to reduce confounding factors, as explained in Section \ref{sec:tasks:metrics}. Another justification is that the raw metrics are not as important in the case for probes, as much as the pattern of learning evident in the model layers, and the relative performance among models. In fact, the regression scores for such tasks show identical learning patterns across the layers for the evaluated list of models. When summarizing the results at a higher level, we simply rank the models.

\myparagraph{Preprocessing.} We tokenize the method samples, which removes the comments, tabs, and new line symbols, which transforms the raw source code into suitable input code representation for the pre-trained models. The method samples are truncated if the maximum sequence length of the model is exceeded, even though it is not a common phenonmenon in the samples of our probing datasets. 

\myparagraph{Collecting feature data.} We pass each sample through the model(s), and collect the vector representation of the sample at every layer. We do not collect the representation of individual tokens, as the features would be too large. We focus on the features of \textit{summary tokens}, e.g., models that follow the \texttt{BERT} architecture have a special \texttt{[CLS]} token at the beginning of the sequence that is used as the overall representation of the entire input \cite{devlin2019bert}. We follow the documentation of the models on the HuggingFace model hub or the respective papers in order to find the correct summary token if they do not have a \texttt{BERT}-style architecture.

\myparagraph{Training.} For training the linear probing classifiers, we train with a batch size of 1 for 20 epochs; we utilize early stopping with a tenacity of 5; we use the Adam optimizer to adapt the learning rate; and we use l2 regularization---optimizing the value of the coefficient during hyper-parameter tuning. \RCtext{R3.7}{The classifier itself is a single linear layer, with a classification layer on top, which takes as input the frozen representation; the size of the layer matches the size of the input.} \RCback

Training performance is evaluated on the validation set. Note that the hyper-parameters are simpler and less impactful since the only training that we do concerns linear probing classifier. We repeat this process for each model and each model layer. Finally, we gather evaluation data on the test set, which we use to present the results of this paper.

\subsection{Pre-trained Transformer models}
We evaluate 8 pre-trained source code models (Table~\ref{tab:models}), and compare their performance against a baseline model, \BERT, which is \textbf{not} specifically trained on source code. 

\BERT is primarily pre-trained on the English language (BooksCorpus---\texttt{800M} words, Wikipedia---\texttt{2.5B} words), 
using the following objectives: Masked Language Modeling (\texttt{MLM}, predicting masked tokens), Next Sentence Prediction (\texttt{NSP}, classifying whether two sentences are related). 

In contrast to the source code models evaluated in this study, \BERT is an interesting reference point as a baseline, since we do not expect that it has knowledge of source code. We briefly discuss the source code models in the following paragraphs. 

\subsubsection{Encoders}
\myparagraph{\CodeBERT} \cite{feng2020codebert} is trained unimodally and bimodally: on just source code, and, on source code with comments. The dataset used is CodeSearchNet \cite{husain_codesearchnet_2019}, which has \texttt{2.1M} bimodal and \texttt{6.4M} unimodal data points, in six programming languages: \textit{Java, Python, JavaScript, PHP, Ruby, Go}. 

\noindent
\textit{Training objective.} The \texttt{MLM} objective is used for the bimodal data points, while a Replaced Token Detection (\texttt{RTD}) objective is used on all the data points. The \texttt{RTD} objective is used to classify whether tokens are original or substituted. \CodeBERT uses n-gram models as generators, one for code, one for natural language.

\myparagraph{\CodeBERTa} \cite{DBLP:journals/corr/abs-1910-03771-codeberta} is pre-trained on data from the CodeSearchNet dataset following the \texttt{RoBERTa} training objective \cite{liu2019roberta}. Note that this model is considerably smaller than the other ones, with only 6 layers and 84 million parameters.

\noindent
\textit{Training objective.} \texttt{RoBERTa} keeps \BERT's \texttt{MLM} objective, but makes it dynamic (each training epoch masks different tokens); it also does not utilize \BERT's \texttt{NSP} objective.

\myparagraph{\GraphCodeBERT} \cite{guo2021graphcodebert} is the model that uses the most structural source code information. During pre-training, \GraphCodeBERT takes as input the nodes of the data-flow graph (\texttt{DFG}), in addition to source code and {natural language comments}. It is also trained on CodeSearchNet. 

\noindent
\textit{Training objective.} It uses three pre-training objectives: \texttt{MLM}; \texttt{DFG} edge prediction (attention edges are masked for 20\% of the nodes and should be predicted); and Node Alignment (predicting edges between code tokens and \texttt{DFG} nodes, for 20\% of \texttt{DFG} nodes).

\myparagraph{\JavaBERT} \cite{9680322javabert} is transformer-based source code model trained specifically on \textit{Java} source code ($\sim$\texttt{3M} \textit{Java} source code files). We use the \emph{javabert-base-cased} checkpoint since \textit{Java} is case-sensitive, and it showed better performance overall compared to the other checkpoint. With just 110M parameters it is one of the smallest models that we probe. 

\noindent
\textit{Training objective.} \JavaBERT is based on the \BERT model with the same masked-language-modelling (MLM) training objective. Note that unlike other models, the input is not processed to
be at the level of functions.

\begin{table*}[]
\scriptsize
\centering
\renewcommand{\arraystretch}{1.1}

\resizebox{0.90\textwidth}{!}{%
\begin{tabular}{|l|c|c|c|c|l|}
 \hline
 \textbf{Model} & \textbf{Params.} & \textbf{Layers}  & \textbf{Heads}  & \textbf{H. Dim.} & \textbf{URL}\\ \hline

 CodeReviewer   & 223M & 12 & 12 & 768 &   {\url{https://huggingface.co/microsoft/codereviewer}} \\ \hline 
 
 CodeT5         & 220M & 12 & 12  & 768 &  {\url{https://huggingface.co/Salesforce/codet5-base}} \\ \hline 
 
 PLBART         & 140M &  6 & 12  & 768 &  {\url{https://huggingface.co/uclanlp/plbart-multi\_task-java}}\\ \hline 
 
 GraphCodeBERT  & 125M & 12 & 12  & 768 &  {\url{https://huggingface.co/microsoft/graphcodebert-base}}\\ \hline 
 
 CodeBERT       & 125M & 12 & 12  & 768 &  {\url{https://huggingface.co/microsoft/codebert-base} }\\ \hline 
 
 UniXCoder      & 125M & 12 & 12  & 768 &  {\url{https://huggingface.co/microsoft/unixcoder-base-unimodal}}\\ \hline 
 
 JavaBERT       & 110M & 12 & 12  & 768 &  {\url{https://huggingface.co/CAUKiel/JavaBERT}}\\ \hline 
 
 
 CodeBERTa      &  84M &  6 & 12  & 768 &  {\url{https://huggingface.co/huggingface/CodeBERTa-small-v1}}\\ \hline 

\end{tabular}}
\bstrut
\caption{Probed models at a glance, ordered by size}
\label{tab:models}
\end{table*}

\subsubsection{Encoder-Decoders}
\myparagraph{\PLBART} \cite{DBLP:journals/corr/abs-2103-06333-plbart} is a bidirectional and auto-regressive transformer which is pre-trained on source code and natural language. It is based on the \texttt{BART} model \cite{DBLP:journals/corr/abs-1910-13461-bart}. The authors of the paper have made multiple models available: we use a version of \PLBART that is trained specifically on \textit{Java}.

\noindent
\textit{Training objective.} It is based on the \texttt{BART} model and pre-trained on a denoising objective: a corrupted data point must be reconstructed. Three noise sources were used: masked tokens, deleted tokens, and token infilling.

\myparagraph{\CodeT} \cite{DBLP:journals/corr/abs-2109-00859-codet5} uses the \texttt{T5} architecture \cite{DBLP:journals/corr/abs-1910-10683}, which unifies all tasks as text generation tasks. \CodeT is trained on CodeSearchNET, along with additional data for \textit{C} and \textit{C\#}. 

\noindent
\textit{Training objective.} \CodeT inherits \texttt{T5}'s Masked Span Prediction (\texttt{MSP}) objective, which is similar to \texttt{MLM}, but the mask can hide a sequence of 1 to 5 tokens. \CodeT is even pre-trained on text-to-code and code-to-text generation. 
\CodeT also uses identifier-specific pre-training tasks: Identifier Tagging (classifying tokens as identifiers or not), and Masked Identifier Prediction (predicting all the identifiers in a snippet).

\myparagraph{\UnixCoder} \cite{guo2022unixcoder} is an unified encoder-decoder pre-trained model, which is trained in a cross-modal manner leveraging \texttt{AST} and code comments to enhance code representations. We probe the \emph{unimodal} checkpoint {in encoder-only mode} since it showed better results for all the probing tasks. 

\noindent 
\textit{Training objective.} \UnixCoder uses three \texttt{NLP} training objectives: \texttt{MLM}, classical left to right next-token language modelling, and \texttt{MSP}. It also uses two code-specific objectives at the level of the entire code fragment. The entire code fragment is encoded and used for cross modal generation (generating the text comment) and contrastive learning (finding the right text comment among several random comments from the same batch). 
\UnixCoder is pre-trained on flattened \texttt{AST}s, but uses the leaves of the \texttt{AST} (i.e., raw source code) for fine-tuning and inference. In addition, the model utilizes mask attention matrices with prefix adapters to control its behaviour (encoder-only, decoder-only, encoder-decoder).

\myparagraph{\CodeReviewer} \cite{li2022codereviewer} is a model designed to understanding code to to assess logic, functionality, latency and other factors as part of code reviewing activities. The model is an encoder-decoder model based on the \texttt{T5} architecture. It is the largest model we probe with \texttt{223M} parameters.

\noindent
\textit{Training objective.} Distinct from other models, this model is pre-trained on code review diffs, rather than source code only, with four pre-training objectives: Diff Tag Prediction (\texttt{DTP}, was a line added, deleted, or kept), Denoising Code Diff (\texttt{DCD}, generating a code line based on context and diff tag), Denoising Review Comment (\texttt{DRC}, \texttt{MSP} for review comments), and Review Comment Generation (\texttt{RCG}, generate review comment given code and other comments). 


\begin{table*}[]
\renewcommand{\arraystretch}{1.2}

\begin{tabular}{|l?c|c?c?c|c|c?c|c|c?c|c|c?c|c|c|}
\hline

  & { \tiny KTX}  & { \tiny IDN}  & { \tiny LEN}  & { \tiny TYP}  & { \tiny REA}  & { \tiny JBL}  & { \tiny SRI}  & { \tiny SRK}  & { \tiny SCK}  & { \tiny OCU}  & { \tiny VCU}  & { \tiny CSC}  & { \tiny MXN}  & { \tiny CPX}  & { \tiny NPT}  \\
\hline
{ \tiny Random }  & {\cellcolor{gray!35} 10}  & {\cellcolor{gray!35} 25}  & {\cellcolor{gray!35} 20}  & {\cellcolor{gray!35} 50}  & {\cellcolor{gray!35} 50}  & {\cellcolor{gray!35} 50}  & {\cellcolor{gray!35} 50}  & {\cellcolor{gray!35} 50}  & {\cellcolor{gray!35} 50}  & {\cellcolor{gray!35} 10}  & {\cellcolor{gray!35} 10}  & {\cellcolor{gray!35} 10}  & {\cellcolor{gray!35} 20}  & {\cellcolor{gray!35} 10}  & {\cellcolor{gray!35} 10}  \\
\hline

{ \tiny BERT} &   {\cellcolor[RGB]{205,229,0} 54.0 } &   {\cellcolor[RGB]{164,229,0} 67.8 } &   {\cellcolor[RGB]{89,229,0} 81.3 } &   {\cellcolor[RGB]{77,229,0} 89.9 } &   {\cellcolor[RGB]{229,193,0} 65.3 } &   {\cellcolor[RGB]{229,87,0} 51.4 } &   {\cellcolor[RGB]{229,112,0} 54.6 } &   {\cellcolor[RGB]{229,171,0} 62.4 } &   {\cellcolor[RGB]{229,164,0} 61.5 } &   {\cellcolor[RGB]{229,119,0} 20.4 } &   {\cellcolor[RGB]{229,127,0} 22.5 } &   {\cellcolor[RGB]{229,168,0} 31.8 } &   {\cellcolor[RGB]{193,229,0} 59.5 } &   {\cellcolor[RGB]{229,181,0} 34.6 } &   {\cellcolor[RGB]{229,157,0} 29.0 } \\ \hline

{ \tiny CodeBERT} &   {\cellcolor[RGB]{119,229,0} 71.8 } &   {\cellcolor[RGB]{104,229,0} 79.5 } &   {\cellcolor[RGB]{82,229,0} 82.8 } &   {\cellcolor[RGB]{35,229,0} 95.3 } &   {\cellcolor[RGB]{126,229,0} 83.5 } &   {\cellcolor[RGB]{220,229,0} 71.2 } &   {\cellcolor[RGB]{225,229,0} 70.5 } &   {\cellcolor[RGB]{229,164,0} 61.5 } &   {\cellcolor[RGB]{229,173,0} 62.7 } &   {\cellcolor[RGB]{229,135,0} 24.0 } &   {\cellcolor[RGB]{229,165,0} 31.1 } &   {\cellcolor[RGB]{229,199,0} 39.0 } &   {\cellcolor[RGB]{168,229,0} 64.7 } &   {\cellcolor[RGB]{229,201,0} 39.5 } &   {\cellcolor[RGB]{229,171,0} 32.3 } \\

{ \tiny CodeBERTa} &   {\cellcolor[RGB]{113,229,0} 73.4 } &   {\cellcolor[RGB]{137,229,0} 73.0 } &   {\cellcolor[RGB]{79,229,0} 83.4 } &   {\cellcolor[RGB]{48,229,0} 93.7 } &   {\cellcolor[RGB]{184,229,0} 75.8 } &   {\cellcolor[RGB]{229,104,0} 53.6 } &   {\cellcolor[RGB]{229,139,0} 58.3 } &   {\cellcolor[RGB]{229,149,0} 59.6 } &   {\cellcolor[RGB]{229,162,0} 61.2 } &   {\cellcolor[RGB]{229,110,0} 18.5 } &   {\cellcolor[RGB]{229,127,0} 23.2 } &   {\cellcolor[RGB]{229,167,0} 31.5 } &   {\cellcolor[RGB]{229,198,0} 45.6 } &   {\cellcolor[RGB]{229,157,0} 29.0 } &   {\cellcolor[RGB]{229,158,0} 29.3 } \\

{ \tiny CReviewer} &   {\cellcolor[RGB]{88,229,0} 79.2 } &   {\cellcolor[RGB]{105,229,0} 79.4 } &   {\cellcolor[RGB]{55,229,0} 88.5 } &   {\cellcolor[RGB]{83,229,0} 89.1 } &   {\cellcolor[RGB]{182,229,0} 76.2 } &   {\cellcolor[RGB]{229,96,0} 52.6 } &   {\cellcolor[RGB]{229,109,0} 54.2 } &   {\cellcolor[RGB]{229,167,0} 61.9 } &   {\cellcolor[RGB]{229,149,0} 59.5 } &   {\cellcolor[RGB]{229,110,0} 19.3 } &   {\cellcolor[RGB]{229,127,0} 22.4 } &   {\cellcolor[RGB]{229,158,0} 29.2 } &   {\cellcolor[RGB]{229,224,0} 51.0 } &   {\cellcolor[RGB]{229,153,0} 28.1 } &   {\cellcolor[RGB]{229,158,0} 29.3 } \\

{ \tiny CodeT5} &   {\cellcolor[RGB]{124,229,0} 70.8 } &   {\cellcolor[RGB]{125,229,0} 75.3 } &   {\cellcolor[RGB]{65,229,0} 86.3 } &   {\cellcolor[RGB]{47,229,0} 93.8 } &   {\cellcolor[RGB]{152,229,0} 80.0 } &   {\cellcolor[RGB]{229,112,0} 54.8 } &   {\cellcolor[RGB]{229,169,0} 62.2 } &   {\cellcolor[RGB]{229,185,0} 64.2 } &   {\cellcolor[RGB]{229,191,0} 65.0 } &   {\cellcolor[RGB]{229,115,0} 20.5 } &   {\cellcolor[RGB]{229,135,0} 24.1 } &   {\cellcolor[RGB]{229,171,0} 32.5 } &   {\cellcolor[RGB]{203,229,0} 57.4 } &   {\cellcolor[RGB]{229,166,0} 31.1 } &   {\cellcolor[RGB]{229,167,0} 31.5 } \\

{ \tiny GCodeBERT} &   {\cellcolor[RGB]{143,229,0} 66.2 } &   {\cellcolor[RGB]{107,229,0} 78.8 } &   {\cellcolor[RGB]{95,229,0} 80.0 } &   {\cellcolor[RGB]{31,229,0} 95.9 } &   {\cellcolor[RGB]{146,229,0} 80.8 } &   {\cellcolor[RGB]{229,224,0} 69.4 } &   {\cellcolor[RGB]{221,229,0} 71.0 } &   {\cellcolor[RGB]{229,186,0} 64.3 } &   {\cellcolor[RGB]{229,167,0} 61.9 } &   {\cellcolor[RGB]{229,134,0} 23.9 } &   {\cellcolor[RGB]{229,177,0} 33.7 } &   {\cellcolor[RGB]{229,201,0} 39.5 } &   {\cellcolor[RGB]{174,229,0} 63.4 } &   {\cellcolor[RGB]{229,190,0} 36.8 } &   {\cellcolor[RGB]{229,180,0} 35.4 } \\

{ \tiny JavaBERT} &   {\cellcolor[RGB]{166,229,0} 60.8 } &   {\cellcolor[RGB]{118,229,0} 76.8 } &   {\cellcolor[RGB]{118,229,0} 75.2 } &   {\cellcolor[RGB]{85,229,0} 88.8 } &   {\cellcolor[RGB]{229,196,0} 65.7 } &   {\cellcolor[RGB]{229,98,0} 52.9 } &   {\cellcolor[RGB]{229,143,0} 58.8 } &   {\cellcolor[RGB]{229,141,0} 58.5 } &   {\cellcolor[RGB]{229,144,0} 59.0 } &   {\cellcolor[RGB]{229,118,0} 19.6 } &   {\cellcolor[RGB]{229,121,0} 20.5 } &   {\cellcolor[RGB]{229,156,0} 28.9 } &   {\cellcolor[RGB]{229,192,0} 44.3 } &   {\cellcolor[RGB]{229,153,0} 28.1 } &   {\cellcolor[RGB]{229,140,0} 25.0 } \\

{ \tiny PLBART} &   {\cellcolor[RGB]{116,229,0} 72.7 } &   {\cellcolor[RGB]{119,229,0} 76.5 } &   {\cellcolor[RGB]{63,229,0} 86.7 } &   {\cellcolor[RGB]{81,229,0} 89.3 } &   {\cellcolor[RGB]{229,169,0} 62.1 } &   {\cellcolor[RGB]{229,77,0} 50.1 } &   {\cellcolor[RGB]{229,80,0} 50.5 } &   {\cellcolor[RGB]{229,144,0} 59.0 } &   {\cellcolor[RGB]{229,139,0} 58.3 } &   {\cellcolor[RGB]{229,100,0} 15.1 } &   {\cellcolor[RGB]{229,109,0} 18.0 } &   {\cellcolor[RGB]{229,147,0} 26.8 } &   {\cellcolor[RGB]{229,156,0} 36.8 } &   {\cellcolor[RGB]{229,140,0} 25.1 } &   {\cellcolor[RGB]{229,139,0} 24.8 } \\

{ \tiny UniXCoder} &   {\cellcolor[RGB]{112,229,0} 73.5 } &   {\cellcolor[RGB]{124,229,0} 75.6 } &   {\cellcolor[RGB]{87,229,0} 81.8 } &   {\cellcolor[RGB]{70,229,0} 90.8 } &   {\cellcolor[RGB]{229,190,0} 65.0 } &   {\cellcolor[RGB]{229,98,0} 52.9 } &   {\cellcolor[RGB]{229,199,0} 66.1 } &   {\cellcolor[RGB]{229,98,0} 52.9 } &   {\cellcolor[RGB]{229,129,0} 56.9 } &   {\cellcolor[RGB]{229,123,0} 20.9 } &   {\cellcolor[RGB]{229,152,0} 27.8 } &   {\cellcolor[RGB]{229,163,0} 30.6 } &   {\cellcolor[RGB]{229,226,0} 51.4 } &   {\cellcolor[RGB]{229,153,0} 28.1 } &   {\cellcolor[RGB]{229,158,0} 29.3 } \\

\hline
\hline
{ \tiny Maximum (10k) } & 
{\cellcolor[RGB]{88,229,0} 79.2 } & 
{\cellcolor[RGB]{104,229,0} 79.5 } & 
{\cellcolor[RGB]{55,229,0} 88.5 } & 

{\cellcolor[RGB]{31,229,0} 95.9 } &  
{\cellcolor[RGB]{126,229,0} 83.5 } &  
{\cellcolor[RGB]{220,229,0} 71.2 } & 

{\cellcolor[RGB]{221,229,0} 71.0 } & 
{\cellcolor[RGB]{229,186,0} 64.3 } &  
{\cellcolor[RGB]{229,191,0} 65.0 } &  

{\cellcolor[RGB]{229,135,0} 24.0 } & 
{\cellcolor[RGB]{229,177,0} 33.7 } & 
{\cellcolor[RGB]{229,201,0} 39.5 } & 

{\cellcolor[RGB]{168,229,0} 64.7 } & 
{\cellcolor[RGB]{229,201,0} 39.5 } & 
{\cellcolor[RGB]{229,180,0} 35.4 } \\

\cline{2-16}
{ \tiny Rank } & 
5\degree & 
4\degree & 
2\degree & 

1\degree & 
3\degree & 
6\degree & 

7\degree & 
10\degree & 
8\degree & 

15\degree & 
14\degree & 
11\degree & 

9\degree & 
11\degree & 
13\degree \\

\hline
\hline
{ \tiny  Std. dev. (exl. BERT) } &






{\cellcolor[RGB]{132,229,0} 5.1 }  & 
{\cellcolor[RGB]{0,229,0} 2.1 }  & 
{\cellcolor[RGB]{81,229,0} 4.0 }  & 
{\cellcolor[RGB]{26,229,0} 2.7 }  & 
{\cellcolor[RGB]{229,213,0} 7.7 }  & 
{\cellcolor[RGB]{229,212,0} 7.7 }  & 
{\cellcolor[RGB]{212,229,0} 6.9 }  & 
{\cellcolor[RGB]{58,229,0} 3.4 }  & 
{\cellcolor[RGB]{14,229,0} 2.5 }  & 
{\cellcolor[RGB]{26,229,0} 2.7 }  & 
{\cellcolor[RGB]{130,229,0} 5.0 }  & 
{\cellcolor[RGB]{98,229,0} 4.4 }  & 
{\cellcolor[RGB]{229,153,0} 9.0 }  & 
{\cellcolor[RGB]{109,229,0} 4.6 }  & 
{\cellcolor[RGB]{62,229,0} 3.3 }  \\

\cline{2-16}
{ \tiny Rank } & 
11\degree & 
1\degree & 
7\degree & 

4\degree & 
13\degree & 
14\degree & 

12\degree & 
6\degree & 
2\degree & 

3\degree & 
10\degree & 
8\degree & 

15\degree & 
9\degree & 
5\degree \\

\hline
\hline
{ \tiny $\Delta$ \xspace with BERT } & 
{\cellcolor[RGB]{0,229,0} 25.2 }  & 
{\cellcolor[RGB]{221,229,0} 11.7 }  & 
{\cellcolor[RGB]{229,163,0} 7.2 }  & 
{\cellcolor[RGB]{229,143,0} 6.0 }  & 
{\cellcolor[RGB]{100,229,0} 18.2 }  & 
{\cellcolor[RGB]{88,229,0} 19.8 }  & 
{\cellcolor[RGB]{144,229,0} 16.4 }  & 
{\cellcolor[RGB]{229,76,0} 1.9 }  & 
{\cellcolor[RGB]{229,102,0} 3.5 }  & 
{\cellcolor[RGB]{229,103,0} 3.6 }  & 
{\cellcolor[RGB]{229,229,0} 11.2 }  & 
{\cellcolor[RGB]{229,171,0} 7.7 }  & 
{\cellcolor[RGB]{229,130,0} 5.2 }  & 
{\cellcolor[RGB]{229,125,0} 4.9 }  & 
{\cellcolor[RGB]{229,150,0} 6.4 }  \\

\cline{2-16}
{ \tiny Rank } & 
1\degree & 
5\degree & 
8\degree & 

10\degree & 
3\degree & 
2\degree & 

4\degree & 
15\degree & 
14\degree & 

13\degree & 
6\degree & 
7\degree & 

11\degree & 
12\degree & 
9\degree \\



\hline
\end{tabular}
\tstrut
\caption{Top: Overall accuracy scores across all tasks. The closer the score is to 100 the more green, the closer it is to random accuracy the more red it is. Bottom: Best performance per task (and rank); Standard deviation of code models (and rank); Delta between best model and \BERT (and rank).}
\label{tab:accuracy}
\end{table*}

\begin{table*}[]
\renewcommand{\arraystretch}{1.2}
\resizebox{\textwidth}{!}{%
\begin{tabular}{|l?r|r?r?r|r|r?r|r|r?r|r|r?r|r|r|}
\hline

  & { \tiny KTX}  & { \tiny IDN}  & { \tiny LEN}  & { \tiny TYP}  & { \tiny REA}  & { \tiny JBL}  & { \tiny SRI}  & { \tiny SRK}  & { \tiny SCK}  & { \tiny OCU}  & { \tiny VCU}  & { \tiny CSC}  & { \tiny MXN}  & { \tiny CPX}  & { \tiny NPT}  \\
\hline
{ \tiny BERT}  & {\cellcolor{blue!35} 0.0 }   & {\cellcolor{blue!35} 0.0 }   & {\cellcolor{blue!35} 0.0 }   & {\cellcolor{blue!35} 0.0 }   & {\cellcolor{blue!35} 0.0 }   & {\cellcolor{blue!35} 0.0 }   & {\cellcolor{blue!35} 0.0 }   & {\cellcolor{blue!35} 0.0 }   & {\cellcolor{blue!35} 0.0 }   & {\cellcolor{blue!35} 0.0 }   & {\cellcolor{blue!35} 0.0 }   & {\cellcolor{blue!35} 0.0 }   & {\cellcolor{blue!35} 0.0 }   & {\cellcolor{blue!35} 0.0 }   & {\cellcolor{blue!35} 0.0 }   \\
\hline

CodeBERT & {\cellcolor[RGB]{229,224,0} 38.7 } &   {\cellcolor[RGB]{229,215,0} 36.3 } &   {\cellcolor[RGB]{229,105,0} 08.0 } &   {\cellcolor[RGB]{176,229,0} 53.5 } &   {\cellcolor[RGB]{182,229,0} 52.4 } &   {\cellcolor[RGB]{227,229,0} 40.7 } &   {\cellcolor[RGB]{229,210,0} 35.0 } &   {\cellcolor{gray!35}} &  {\cellcolor[RGB]{229,87,0} 3.1 } &   {\cellcolor[RGB]{229,93,0} 4.4 } &   {\cellcolor[RGB]{229,119,0} 11.1 } &   {\cellcolor[RGB]{229,116,0} 10.6 } &   {\cellcolor[RGB]{229,126,0} 12.8 } &   {\cellcolor[RGB]{229,105,0} 7.5 } &   {\cellcolor[RGB]{229,94,0} 4.6 }  \\

CodeBERTa & {\cellcolor[RGB]{221,229,0} 42.2 } &   {\cellcolor[RGB]{229,138,0} 16.1 } &   {\cellcolor[RGB]{229,118,0} 11.2 } &   {\cellcolor[RGB]{229,218,0} 37.6 } &   {\cellcolor[RGB]{229,192,0} 30.3 } &   {\cellcolor[RGB]{229,93,0} 4.5 } &   {\cellcolor[RGB]{229,107,0} 8.1 } &   {\cellcolor{gray!35}} &  {\cellcolor[RGB]{229,80,0} 1.0 } &  {\cellcolor{gray!35}} &  {\cellcolor{gray!35}} &  {\cellcolor{gray!35}} &  {\cellcolor{gray!35}} &  {\cellcolor{gray!35}} &  {\cellcolor[RGB]{229,79,0} 0.4 }  \\

CReviewer & {\cellcolor[RGB]{173,229,0} 54.8 } &   {\cellcolor[RGB]{229,214,0} 36.0 } &   {\cellcolor[RGB]{229,222,0} 38.5 } &   {\cellcolor{gray!35}} &  {\cellcolor[RGB]{229,195,0} 31.4 } &   {\cellcolor[RGB]{229,86,0} 2.5 } &   {\cellcolor{gray!35}} &  {\cellcolor{gray!35}} &  {\cellcolor{gray!35}} &  {\cellcolor{gray!35}} &  {\cellcolor{gray!35}} &  {\cellcolor{gray!35}} &  {\cellcolor{gray!35}} &  {\cellcolor{gray!35}} &  {\cellcolor[RGB]{229,79,0} 0.4 }  \\

CodeT5 & {\cellcolor[RGB]{229,216,0} 36.5 } &   {\cellcolor[RGB]{229,166,0} 23.3 } &   {\cellcolor[RGB]{229,178,0} 26.7 } &   {\cellcolor[RGB]{229,226,0} 38.6 } &   {\cellcolor[RGB]{220,229,0} 42.4 } &   {\cellcolor[RGB]{229,102,0} 7.0 } &   {\cellcolor[RGB]{229,140,0} 16.7 } &   {\cellcolor[RGB]{229,94,0} 4.8 } &   {\cellcolor[RGB]{229,111,0} 9.1 } &   {\cellcolor[RGB]{229,78,0} 0.1 } &  {\cellcolor[RGB]{229,85,0} 2.1 } &   {\cellcolor[RGB]{229,80,0} 1.0 } &   {\cellcolor{gray!35}} &  {\cellcolor{gray!35}} &  {\cellcolor[RGB]{229,89,0} 3.5 }   \\

GCodeBERT & {\cellcolor[RGB]{229,177,0} 26.5 } &   {\cellcolor[RGB]{229,208,0} 34.2 } &   {\cellcolor{gray!35}} &  {\cellcolor[RGB]{155,229,0} 59.4 } &   {\cellcolor[RGB]{211,229,0} 44.7 } &   {\cellcolor[RGB]{229,218,0} 37.0 } &   {\cellcolor[RGB]{229,214,0} 36.1 } &   {\cellcolor[RGB]{229,96,0} 5.1 } &   {\cellcolor[RGB]{229,79,0} 1.0 } &   {\cellcolor[RGB]{229,93,0} 4.3 } &   {\cellcolor[RGB]{229,133,0} 14.4 } &   {\cellcolor[RGB]{229,119,0} 11.3 } &   {\cellcolor[RGB]{229,113,0} 9.6 } &   {\cellcolor[RGB]{229,89,0} 3.4 } &   {\cellcolor[RGB]{229,111,0} 9.0 }    \\

JavaBERT & {\cellcolor[RGB]{229,132,0} 14.8 } &   {\cellcolor[RGB]{229,183,0} 28.0 } &   {\cellcolor{gray!35}} &  {\cellcolor{gray!35}} &  {\cellcolor[RGB]{229,79,0} 1.2 } &   {\cellcolor[RGB]{229,88,0} 3.1 } &   {\cellcolor[RGB]{229,111,0} 9.3 } &   {\cellcolor{gray!35}} &  {\cellcolor{gray!35}} &  {\cellcolor{gray!35}} &  {\cellcolor{gray!35}} &  {\cellcolor{gray!35}} &  {\cellcolor{gray!35}} &  {\cellcolor{gray!35}} &  {\cellcolor{gray!35}}   \\

PLBART & {\cellcolor[RGB]{227,229,0} 40.7 } &   {\cellcolor[RGB]{229,180,0} 27.0 } &   {\cellcolor[RGB]{229,185,0} 28.9 } &   {\cellcolor{gray!35}} &  {\cellcolor{gray!35}} &  {\cellcolor{gray!35}} &  {\cellcolor{gray!35}} &  {\cellcolor{gray!35}} &  {\cellcolor{gray!35}} &  {\cellcolor{gray!35}} &  {\cellcolor{gray!35}} &  {\cellcolor{gray!35}} &  {\cellcolor{gray!35}} &  {\cellcolor{gray!35}} &  {\cellcolor{gray!35}}   \\

UniXCoder & {\cellcolor[RGB]{220,229,0} 42.4 } &   {\cellcolor[RGB]{229,169,0} 24.2 } &   {\cellcolor[RGB]{229,85,0} 2.7 } &   {\cellcolor[RGB]{229,110,0} 8.9 } &   {\cellcolor{gray!35}} &  {\cellcolor[RGB]{229,88,0} 3.1 } &   {\cellcolor[RGB]{229,173,0} 25.3 } &   {\cellcolor{gray!35}} &  {\cellcolor{gray!35}} &  {\cellcolor[RGB]{229,79,0} 0.5 } &   {\cellcolor[RGB]{229,98,0} 6.8 } &   {\cellcolor{gray!35}} &  {\cellcolor{gray!35}} &  {\cellcolor{gray!35}} &  {\cellcolor[RGB]{229,79,0} 0.4 }   \\

\hline
\end{tabular}}
\tstrut
\caption{Comparison with the \BERT baseline (normalized w.r.t the \BERT accuracy as the lower limit and 100 as the upper limit for each task; cells are marked grey where accuracy is below the \BERT baseline accuracy).} 
\label{tab:bert_baseline}
\end{table*}

\section{Results}
\label{sec:results}

In this section, we discuss the probing results, in 3 parts: Task analysis, Model Analysis, and Layer Analysis.

Due to the large number of tasks, we often group them into the categories that they were presented in, in Section~\ref{sec:tasks}.
The results in this section refer to Tables~\ref{tab:accuracy} and~\ref{tab:bert_baseline} below.

\subsection{Task analysis} 

\noindent
To determine how effective models are against the probes, we first take a look at their collective performance, and pose the following research questions below. This gives us an idea as to which characteristics for which tasks, are well-encoded in the hidden layers of the models.

\begin{itemize}
   \item \RQ{1.1} To what extent can current source code models encode intrinsic code characteristics?
   \item \RQ{1.2} To what extent do source code models encode code characteristics \emph{better than} the baseline?
\end{itemize}

To answer \RQ{1.1}, we look at the maximum model performance for each task, as well as the standard deviation of source code model accuracies for each task. 

To answer \RQ{1.2}, we compare the difference between highest accuracy recorded for a task and the \BERT baseline accuracy, which is not trained on source code.

\subsubsection{Collective performance on tasks}

\vskip 1mm
\textbf{\tokbased tasks:} For \tokbased tasks such as \KTX (79.2\%), \IDN (79.5\%), as well as \LEN (88.5\%), we observe good overall performance. In terms of the standard deviation among the source code models, \IDN and \LEN have low variance ($sigma$ ranks 4 and 2), with \KTX exhibiting somewhat higher variance ($sigma$ rank 11).

Therefore, since the accuracies of the models are generally high, with a generally low variance in scores, this shows that identifier-based syntactic information and token-level surface information are well-encoded in all the models, although information pertaining to keywords is more model-dependent. 

\vskip 1mm
\textbf{\bugbased tasks:} For \bugbased tasks such \TYP (95.9 \%), \REA (83.5\%), and \JBL (71.2\%), we see considerable variations in performance. For the \TYP task, all models perform considerably well, with a low variance across the model scores ($sigma$ rank 4). In fact, if we look at the layer-wise performance of each model, almost all layers reflect high scores, indicating this is an easy task; and intrinsic information relevant to identifying \emph{unnatural} misspelled types are well-encoded in the model hidden layers. 

For \REA even though the best reported score is high (83\%) the variance among the model scores is the highest ($\sigma$ rank 13) For \JBL, the variance is slightly higher ($\sigma$ rank 14), while the performance is somewhat lower (71.2\%). This indicates that although some models may perform well or moderately well in term of accuracy, other models may struggle to make good predictions altogether. Indeed, the lowest scores for this task range from 50-60\%, which is clearly not impressive considering that the random accuracy for these tasks is 50\% (binary classification). This shows us that there is room for improvement for some models in encoding comprehensive semantic information relevant to these tasks.

\vskip 1mm
\textbf{\switchbased tasks:} \switchbased semantic tasks such as \SRI (71 \%), \SRK (64.3\%), and \SCK (65\%), report modest prediction accuracies. Moreover, the variance for the best-performing task, \SRI, is high ($\sigma$ rank 12). This indicates that while some of the models may have significant awareness of this type of semantic information, others struggle. In fact, several models have performance below 60\% (some close to 50\%), which is very low for a binary classification task with a random chance accuracy of 50\%. 

The more difficult tasks \SRK and \SCK start to show the limits of current source code models, as it is more challenging for them to distinguish these more difficult cases. While performance is relatively low, the variance is also low ($sigma$ ranks 6 and 2), indicating that the models have more comparable performance. Overall, this indicates that there is significant room for improvement for this task category.

\vskip 1mm
\textbf{\countbased tasks:} For \countbased tasks such as \OCU (24.0\%), \VCU (33.7\%), and \CSC (39.5\%), we see evidence that source code models struggle with surface-level and structural information. This is particularly for \OCU, which is the worst-performing task overall. While the models score significantly better than the naive accuracy (10\%), the overall accuracies demonstrate that these are harder tasks. For comparison, we see that \KTX has a similar naive accuracy, but the models perform much better. The variance between models is intermediate ($sigma$ ranks: 3, 10, and 8), indicating that some models tend to struggle less than others.

\vskip 1mm
\textbf{\compbased tasks:} Finally, the \compbased tasks show further evidence that source code models struggle with structural information. While \MXN has relatively good performance (64.7\%), \CPX and \NPT are on par with the \countbased tasks (respectively 39.5\% and 35.4\%). In terms of variance, the models are very spread out for \MXN (the highest of all tasks), and moderate for \CPX and \NPT ($sigma$ ranks 9 and 5). This shows that some models struggle more than others to encode structural information relating to the control-flow in code (particularly regarding indentation).

\subsubsection{Comparison with the baseline}

Comparing the performance of the models with \BERT accentuates the emerging trend in \RQ{1.1}. While for some tasks, the performance increase over BERT is very high (as high as 25\%), for others, the source code models offer very little improvements (as low as 2\%). In terms of task categories, \tokbased and \switchbased tasks are the ones where the largest improvements are seen, although there is variation accross tasks. \switchbased and \countbased tasks exhibit moderate improvements as a group (again with some variation), with \compbased tasks having the smallest improvements as a group.

Given the performance variation as a group, we briefly discuss the tasks with the highest and lowest improvements. The tasks with the largest improvement are \KTX, \JBL, and \REA (tasks related to the semantic of individual keywords and of the method), followed by \SRI, \IDN, and \VCU (all tasks that have to do with identifiers). On the other hand, the tasks with the smallest improvements are \SRK, \SCK, \OCU, \CPX, and \MXN. These include two of the \compbased tasks, as well tasks relating to keywords in context, and operators. These tasks are also tasks where the models are struggling in the first place.

\vskip 1mm %
\noindent
\subsubsection{Summary} 

The overall picture that emerges is that source code models can do very well on {syntactic} token tagging tasks, and tasks that probe on the {surface}-level token-length, and on a subset of our {semantic} tasks such as \bugbased tasks. 

For harder semantic \switchbased tasks, the models perform moderately when identifiers are involved (\SRI), and less well when keywords are involved. This shows that there is room for improvement for comprehensive semantic understanding in models. Furthermore, we observe that there are clear issues with tasks that involve more {structural} knowledge of source code, that is less easily extracted from a sequence of tokens.

Finally, a further sign that models are struggling is that for the more difficult tasks, the improvements over tend to be small overall.

\resultbox{\textit{Source code models tend to perform well on tasks that probe syntactic and some {semantic} characteristics, but they clearly struggle with tasks that probe more {structural} characteristics.}}


\def\minigraphx#1#2#3{
  \def\freqcolor{#1}
  \begin{tikzpicture}[xscale=0.08, yscale=0.08]
    \useasboundingbox (-0.2,0.6) rectangle (#2.2,4.2); 
    \begin{scope}[ycomb, yscale=0.3]
      \draw[fill=\freqcolor!100!black!10, \freqcolor!100!black!10] (0,0) rectangle (#2,10); 
      \draw[xshift=-14pt,  \freqcolor!70!black!80, line width=1.15] plot[] file {#3}; 
      \draw[\freqcolor!100!black!25, line width=0.2] (-0.1,-0.4) rectangle (#2.1,10.4);  
   \end{scope}
\end{tikzpicture}
}

\def\minigraphxtext#1#2#3{
  \def\freqcolor{#1}
  \begin{tikzpicture}[xscale=0.08, yscale=0.08]
    \begin{scope}[ycomb, yscale=0.3]
      \draw[xshift=-14pt,  \freqcolor!70!black!80, line width=1.15] plot[] file {#3}; 
      \draw[\freqcolor!100!black!25, line width=0.1] (-0.1,-0.1) rectangle (12.0,0.12);  
   \end{scope}
\end{tikzpicture}
}

\def\minigraphmax#1#2#3#4{
  \def\freqcolor{#1}
  \begin{tikzpicture}[xscale=0.08, yscale=0.08]
    \useasboundingbox (-0.2,1.9) rectangle (#2.2,4.2); 
    \begin{scope}[ycomb, yscale=0.3]
      \draw[fill=\freqcolor!100!black!10, \freqcolor!100!black!10] (0,0) rectangle (#2,#4); 
      \draw[xshift=-14pt,  \freqcolor!70!black!80, line width=1.15] plot[] file {#3}; 
   \end{scope}
\end{tikzpicture}
}

\def\minigraphmaxtext#1#2#3#4{
  \def\freqcolor{#1}
  \begin{tikzpicture}[xscale=0.08, yscale=0.08]
    \begin{scope}[ycomb, yscale=0.3]
      \draw[xshift=-14pt,  \freqcolor!70!black!80, line width=0.9] plot[] file {#3}; 
      \draw[\freqcolor!100!black!25, line width=0.1] (-0.1,-0.1) rectangle (#2.0,0.#2);  
   \end{scope}
\end{tikzpicture}
}

\def\fullstats{
\vocabstats{full}{11.6M, 100}{2.43B, 100}{42}{79}{83}
}

\def\minigraph#1#2#3{

\minigraphx{#1}{#2}{foo}
}

\def\modelperf#1{
\minigraphx{orange}{6}{#1-syntactic} \minigraphx{blue}{5}{#1-semantic} \minigraphx{green}{8}{#1-structural} \minigraphx{red}{4}{#1-surface}
}


\def\modelperfx#1{
Syn: \minigraphx{orange}{6}{#1-syntactic}\emoji{1st-place-medal} Sem: \minigraphx{blue}{5}{#1-semantic}\emoji{3rd-place-medal} Struc:\minigraphx{green}{8}{#1-structural}\emoji{black-circle} Surf: \minigraphx{red}{4}{#1-surface}\emoji{red-circle}
}

\def\alltasksgraph#1{
  \def \ntasks{20}
  \def\bgcolor{lightgray}
   \def\tokencolor{blue}
   \def\lencolor{black}
   \def\metriccolor{orange}
   \def\complexcolor{red}
   \def\debugcolor{green}
   \def\switchcolor{cyan}
   \def\linecolor{lightgray}
  \begin{tikzpicture}[xscale=0.08, yscale=0.08]
    \useasboundingbox (-0.2,0.6) rectangle (\ntasks.2,4.2); 
    \begin{scope}[ycomb, yscale=0.3]
      \draw[fill=\bgcolor!100!black!10, \bgcolor!100!black!10] (0,0) rectangle (\ntasks,10); 
      \draw[xshift=-14pt,  \tokencolor!70!black!80, line width=1.15] plot[] file {#1-token}; 
      \draw[xshift=-14pt,  \lencolor!70!black!80, line width=1.15] plot[] file {#1-len};
      \draw[xshift=-14pt,  \metriccolor!70!black!80, line width=1.15] plot[] file {#1-metrics};
      \draw[xshift=-14pt,  \complexcolor!70!black!80, line width=1.15] plot[] file {#1-complex};
      \draw[xshift=-14pt,  \debugcolor!70!black!80, line width=1.15] plot[] file {#1-debug};
      \draw[xshift=-14pt,  \switchcolor!70!black!80, line width=1.15] plot[] file {#1-switch};
      \draw[\linecolor!100!black!25, line width=0.1] (-0.1,-0.4) rectangle (\ntasks.1,10.4);  
   \end{scope}
\end{tikzpicture}
}

\def\taskperf2#1{
T: \minigraphx{blue}{2}{#1-token} M: \minigraphx{orange}{4}{#1-metrics} \minigraphx{red}{3}{#1-complex} S: \minigraphx{green}{3}{#1-debug} \minigraphx{cyan}{3}{#1-switch}
}

\def\taskperf#1{
T\minigraphx{blue}{2}{#1-token} M\minigraphx{orange}{4}{#1-metrics} C\minigraphx{red}{3}{#1-complex} I\minigraphx{green}{3}{#1-debug}S\minigraphx{cyan}{3}{#1-switch}
}

\def\layerperfbig#1{
\minigraphmax{green}{12}{layers-#1}{12}
}

\def\layerperfsmall#1{
\minigraphmax{green}{6}{layers-#1}{12}
}

\def\layerperfbigtext#1{
\minigraphmaxtext{green}{12}{layers-#1}{12}
}

\def\layerperfsmalltext#1{
\minigraphmaxtext{green}{6}{layers-#1}{12}
}

\def\tasklayerperf#1{
\minigraphxtext{blue}{12}{task-#1}
}

\def\alltasksbertmax#1{
  \def \ntasks{20}
  \def\bgcolor{lightgray}
   \def\tokencolor{blue}
   \def\lencolor{blue}
   \def\metriccolor{orange}
   \def\complexcolor{red}
   \def\debugcolor{green}
   \def\switchcolor{cyan}
   \def\linecolor{lightgray}


  \begin{tikzpicture}[xscale=0.08, yscale=0.08]
    \useasboundingbox (7.8,1.8) rectangle (\ntasks.2,4.2); 
    \begin{scope}[ycomb, yscale=0.3]
      \draw[fill=\bgcolor!100!black!10, \bgcolor!100!black!10] (0,0) rectangle (15,10); 
      \draw[xshift=-14pt,  \tokencolor!70!black!80, line width=1.15] plot[] file {ident-#1}; 
      \draw[xshift=-14pt,  \lencolor!70!black!80, line width=1.15] plot[] file {len-#1};
      \draw[xshift=-14pt,  \debugcolor!70!black!80, line width=1.15] plot[] file {bugs-#1};
      \draw[xshift=-14pt,  \switchcolor!70!black!80, line width=1.15] plot[] file {switch-#1};
      \draw[xshift=-14pt,  \metriccolor!70!black!80, line width=1.15] plot[] file {metrics-#1};
      \draw[xshift=-14pt,  \complexcolor!70!black!80, line width=1.15] plot[] file {complex-#1};
   \end{scope}
\end{tikzpicture}
}

\subsection{Model analysis}

\noindent
In this section, we inspect the performance of the models individually across tasks, and task categories. We pose the following research questions as to ascertain which models encode the most information across tasks: 

\begin{itemize}
\item \RQ{2.1} Which models are the best in each category?
\item \RQ{2.2} What are the best models overall? 
\end{itemize}

\subsubsection{Performance on task categories.} 
For \RQ{2.1}, we focus on the performance of the models in each task category, and the performance against the \BERT baseline. 
Table \ref{tab:bert_baseline} shows the performance of the source code models, normalized against the \BERT baseline accuracy. 

\textbf{\tokbased tasks:} For token-based tasks such as \KTX and \IDN, all source code models comfortably exceed the \BERT baseline accuracy. For \KTX, \CodeReviewer is the best performing model, while, for \IDN, \CodeBERT and \CodeReviewer report the best probing accuracies. 

The \BERT accuracies for \KTX and \IDN are the lowest among the other evaluated models, indicating that, as expected, source code models are as a whole much more proficient in learning such code-specific characteristics related to keyword and identifier syntax than \BERT. Almost all models outperform \BERT on \KTX by more than 10\%, and all models exceed \BERT's performance on \IDN by at least 5\%, with several exceeding 10\%.

Although \BERT does quite well for the \LEN task, most code models outperform \BERT significantly, suggesting that they have a greater knowledge of surface-level information that are encoded in their hidden layers. Six source code models exceed the \BERT baseline by as much as 7\% for the \LEN task. \CodeReviewer is the best performing model, while \PLBART and \CodeT are competitive. Interestingly, two \BERT-based source code models are under-performing on \LEN (\GraphCodeBERT and \JavaBERT), while the top three models all have an encoder-decoder architecture, 
suggesting that the encoding of intrinsic information relevant to \LEN may be influenced by model architecture.

\resultbox{\textit{We note that for \KTX, \IDN, and \LEN tasks, \CodeReviewer is consistently among the best performing models.}}

\begin{table}
\scriptsize
\centering
\renewcommand{\arraystretch}{2.0}
\tstrut

\resizebox{\linewidth}{!}{%
\begin{tabular}{|l|c|c|c|c|c|c|}
\hline

Model & vs. BERT & Layers & Rank 1 & Rank 2 & Rank 3 & $<$ BERT\\
\hline

\hline
\GraphCodeBERT  & {\cellcolor[RGB]{249, 249, 249}\alltasksbertmax{GCodeBERT}}
                & {\cellcolor[RGB]{230, 255, 230}\layerperfbig{GraphCodeBERT}}  
                & {\gold 6}        
                & {\silver 5}         
                & {\bronze 2} 
                & 1 \\

\hline
\CodeBERT       & {\cellcolor[RGB]{249, 249, 249}\alltasksbertmax{CodeBERT}}
                & {\cellcolor[RGB]{230, 255, 230}\layerperfbig{CodeBERT}}       
                & {\gold 6}        
                & {\silver 6}         
                & ---
                & 1 \\

\hline
\CodeT          & {\cellcolor[RGB]{249, 249, 249}\alltasksbertmax{CodeT5}}
                & {\cellcolor[RGB]{230, 255, 230}\layerperfbig{CodeT5}}         
                & {\gold 1}        
                & {\silver 1}         
                & {\bronze 6} 
                & 2 \\

\hline
\CodeReviewer   & {\cellcolor[RGB]{249, 249, 249}\alltasksbertmax{CReviewer}}
                & {\cellcolor[RGB]{230, 255, 230}\layerperfbig{CodeReviewer}}   
                & {\gold 2}        
                & {\silver 1}         
                & --- 
                & 9 \\

\hline
\UnixCoder      & {\cellcolor[RGB]{249, 249, 249}\alltasksbertmax{UniXCoder}}
                & {\cellcolor[RGB]{230, 255, 230}\layerperfbig{UnixCoder}}      
                & ---        
                & {\silver 1}        
                & {\bronze 3} 
                & 6 \\

\hline
\BERT           & {\cellcolor[RGB]{249, 249, 249} --- }
                & {\cellcolor[RGB]{230, 255, 230}\layerperfbig{BERT}}           
                &  ---      
                &  --- 
                & {\bronze 3} 
                & --- \\

\hline
\PLBART         & {\cellcolor[RGB]{249, 249, 249}\alltasksbertmax{PLBART}}
                & {\cellcolor[RGB]{230, 255, 230}\layerperfbig{PLBART}}         
                & ---        
                & {\silver 1}         
                & ---                            
                & 12 \\

\hline
\CodeBERTa      & {\cellcolor[RGB]{249, 249, 249}\alltasksbertmax{CodeBERTa}}
                & {\cellcolor[RGB]{230, 255, 230}\layerperfbig{CodeBERTa}}      
                &   ---      
                &   ---      
                & {\bronze 1} 
                & 6 \\

\hline
\JavaBERT       & {\cellcolor[RGB]{249, 249, 249}\alltasksbertmax{JavaBERT}}
                & {\cellcolor[RGB]{230, 255, 230}\layerperfbig{JavaBERT}}       
                &   ---    
                & ---         
                &   ---                          
                & 10 \\

\hline
\end{tabular}}
\bstrut
\caption{Model Performance summary: task performance sparklines, layer performance sparklines, and medal tally}
\label{tab:model_tally}
\end{table}

\textbf{\bugbased tasks:} For incorrect code tasks, almost all source code models exceed the \BERT baseline. In all cases, \CodeBERT and \GraphCodeBERT vie for the top two positions. \CodeT is consistently third, with \CodeBERTa and \CodeReviewer being occasionally competitive. Importantly, while all but one models exceed \BERT's performance for \JBL, only \CodeBERT and \GraphCodeBERT do so with a consequent margin (19.8\% and 18\%), outlining that this is a more challenging task. On the other hand, five models outperform \BERT by margin that exceeds 10\% on \REA.

\textbf{\switchbased tasks:} For the semantic-replacement task \SRI, two source code models outperform the \BERT-baseline by at least 15\%: \GraphCodeBERT, \CodeBERT. They are followed by \UnixCoder and \CodeT, which have also sizeable improvements over \BERT. Other models are closer to \BERT, if not worse. For both \SCK and \SRK however, \emph{most models perform worse than \BERT}. Only \CodeT, \GraphCodeBERT, and \CodeBERT can surpass \BERT's performance, and do so by small margins even in the best of cases (less than 2\% for \SRK, 3.5\% for \SCK). Most models are close together (except \UnixCoder on \SRK). If there is a silver lining, it is that the difference is larger on \SCK. Since \SCK requires that models find replacements of compatible rather than random keywords, it is a task that requires more source code knowledge than \SRK. Indeed, \BERT does worse on \SCK than \SRK, while some models such as \CodeT or \CodeBERT do slightly better. \BERT actually ranks \emph{third} on the \SRK task (fourth on \SCK).

\resultbox{\textit{\GraphCodeBERT, \CodeT, and  \CodeBERT are consistently among the best for Mistyped and Replacement tasks.}}

\textbf{\countbased tasks:} For \OCU, \VCU, and \CSC, \emph{most source code models perform worse than \BERT}. Only two models perform consistently well (\GraphCodeBERT and \CodeBERT), with \UnixCoder a distant third. A silver lining is that on \VCU and \CSC, \GraphCodeBERT and \CodeBERT outperform \BERT by noticeable to consequent margins (7 to 11\%). In particular, \GraphCodeBERT does comparatively well on \VCU (+11.7\% over \BERT). One reason for this might be that reasoning about variables is very well aligned with \GraphCodeBERT's original training objective. That said, in absolute terms, \GraphCodeBERT's performance is limited (barely a third of correct predictions).


\textbf{\compbased tasks:} For \compbased tasks, \emph{only two models outperform \BERT for \CPX and \MXN}. In both cases, \CodeBERT is first, followed by \GraphCodeBERT. Thus, while at first glance the performance on \MXN appears to be relatively good at 64.7\%, most source code models are actually far below that level (ranging from 36 to 57\%). Surprisingly, the situation is somewhat better for the more challenging \NPT task: more source code models outperform \BERT. The top three (\GraphCodeBERT, \CodeBERT, and \CodeT) do so by a good margin (particularly for \GraphCodeBERT). Three more models \emph{barely} exceeds the baseline's performance. We were surprised that \BERT ranks \emph{third} in two of the three complexity tasks.

\resultbox{\textit{While \CodeBERT and \GraphCodeBERT perform best, no model performs well for Count and Complexity tasks.}}

\subsubsection{Overall model performance} For \RQ{2.2}, we focus on the ranking of models across all tasks, as presented in Table~\ref{tab:model_tally}. The table summarizes the relative performance amongst the models by counting the number of times the models rank first, second, or third on any given task. We also provide summary of each model's performance via a sparkline inspired graph \cite{tufte2004sparklines}: tasks are ordered by categories (\tokbased in blue, \bugbased in green, \switchbased in cyan, \countbased in orange, \compbased in red). The layers are normalized with the baseline \BERT performance as the lower bound and the maximum performance for the task as 
the upper bound. 

We can see that two models clearly stand out: \CodeBERT and \GraphCodeBERT. They are essentially tied. While \CodeBERT ranks second in more tasks (6 versus 5), \GraphCodeBERT arrives in the top 3 in 13 out of 15 tasks, compared to 12 for \CodeBERT. Both models perform below \BERT only once (on \LEN for \GraphCodeBERT, and on \SRK for \CodeBERT). 

\CodeT is a clear, but distant third: it is placed in the top 3 models in 8 tasks (\SCK, \SRK, \REA, \JBL, \CSC, \TYP, \LEN, and \NPT), while performing worse than \BERT on only \MXN and \CPX. \CodeReviewer is fourth, ranking in the top 3 on 3 tasks (twice first, on \KTX and \LEN, and second on \IDN), but performs worse than \BERT on \emph{9 tasks}. \UnixCoder follows, placing second on \KTX and third on \SRI, \OCU, and \VCU, while faring worse than \BERT on 6 tasks.

The three remaining models (\PLBART, \JavaBERT, and \CodeBERTa) all perform \emph{worse overall than \BERT}. This leaves \BERT as sixth overall, in particular finishing third on \emph{four tasks}: \SRK, and two of the three \compbased tasks (\MXN and\CPX).

\resultbox{\textit{\CodeBERT and \GraphCodeBERT are by far the best performing model, followed by \CodeT. Surprisingly, three source code models perform {worse} than the \BERT baseline.}}


\subsection{Layer analysis}

\noindent
\RCtext{R3.C9}{Studies in the field of NLP have shown for the case of language, that pre-trained models such as BERT tend to represent surface characteristics of language best in early layers, syntactic characteristics in the middle layers, and higher-level semantic characteristics in the later layers} \cite{jawahar-etal-2019-bert}. \RC{These studies contribute to our scientific understanding of these models, which is why we perform a similar study in the case of source code models.
To understand if learning patterns exist across tasks for a given model, or across models for a given task, we analyze the layer-by-layer performance of the models}.\RCback We aggregate the model scores into a single inline graph which provides a glimpse into the overall task learning. Even though the learning patterns may differ from model to model at an individual-level, the collective performance indicators help us understand where relevant characteristics may be encoded. 

For the previous analyses, we always considered the best performing layer for each model. In this section, we inspect the performance of the models at the layer-level. We divide our analysis into two parts, layer analysis by tasks and by models. We pose the following research questions to determine in which layers learning has taken place most effectively for the probed characteristics.

\begin{itemize}
\item \RQ{3.1} {For a given task, which layers show the most effective learning of  
code characteristics?} 
\item \RQ{3.2} Is the final layer universally suitable across tasks?
\end{itemize}

For space and clarity reasons, we simplify the presentation by focusing on trends and patterns, instead of presenting a large amount of numbers corresponding to individual layer-wise accuracies of each model, across fifteen tasks. We present inline graphs inspired by sparklines \cite{tufte2004sparklines} that show the relative performance of each layer. Detailed layer-wise heatmaps of model accuracies for all tasks are available in Appendix \ref{appendix:heatmaps}. For each task and each model, we first rank the layers to homogenize the performance of models and tasks (higher scores are better) and reduce the influence of outliers. The ranks are then averaged over all 12-layer models for a single task \RQ{3.1} (e.g., \LEN: \tasklayerperf{LEN}), or averaged over all tasks for a single model \RQ{3.2} (e.g., \CodeBERT: \layerperfbigtext{CodeBERT}), before being depicted in sparklines.

\subsubsection{Layer performance by task}

Although the very last layer often performs well, and better than its immediate predecessors---leading to a ``last-layer peak'' as shown in e.g., \CSC \tasklayerperf{CSC}---, we note that the performance is clearly \emph{not} always best in the latest layers. Overall, we can group tasks based on whether the best performance arises in early, early-middle, middle, or late layers. 

\myparagraph{Early layers.} Three tasks show aggregate best performance in the early layers, in two distinct fashions. First, \KTX \tasklayerperf{KTX} clearly peaks at the first layer. Second, \LEN \tasklayerperf{LEN} and \TYP \tasklayerperf{TYP} have a broad peak in the first half of layers, with clear lower scores in the last layers.

\myparagraph{Early-Middle layers.} Nine tasks exhibit their highest performance in the middle layers. Of these, four show a comparatively earlier peak, leading to this classification. These are \CSC \tasklayerperf{CSC}, \CPX \tasklayerperf{CPX}, \NPT \tasklayerperf{NPT} and \REA \tasklayerperf{REA}.

\myparagraph{Middle layers.} Three tasks show best scores in the middle layers, with the last layer usually also performing well. These are \VCU \tasklayerperf{VCU}, \SRK \tasklayerperf{SRK}, and \MXN \tasklayerperf{MXN}. Additionally, \OCU \tasklayerperf{OCU} and \IDN \tasklayerperf{IDN} show similar, albeit less-pronounced patterns.

\myparagraph{Later layers.} Finally, three tasks show a trend of almost continuous improvement from the early layers to the final layers. These are \JBL \tasklayerperf{JBL}, \SRI \tasklayerperf{SRI}, and \SCK \tasklayerperf{SCK}.

\myparagraph{Relation to task categories.} We note that two \tokbased tasks peak in early layers (\KTX, \LEN), while \IDN, which relies more on identifier semantics, peaks in the middle layers. For \emph{structural} tasks (\countbased and \compbased), three tasks peak in the early-middle layers (\CSC, \CPX, \NPT), and three others in the middle layers (\VCU, \OCU, and \MXN). Finally, the more \emph{semantic} tasks (\bugbased and \switchbased) have one task peaking in early layers (\TYP---the easiest), one each in the early-middle (\REA) and middle (\SRK) layers, and three tasks peaking in the late layers (\JBL, \SRI, and \SCK). Thus, while we caution against over-interpreting these results, we note a tendency for structural tasks to perform better in the middle layers, and for more semantic tasks to perform better in the later layers, with some tasks that do not fulfill this model.

\subsubsection{Layer performance by model}

For the models, we also see several patterns of performance across layers.

\myparagraph{Middle layers.} \JavaBERT \layerperfbigtext{JavaBERT}, \GraphCodeBERT \layerperfbigtext{GraphCodeBERT}, and to a lesser extent \CodeBERT \layerperfbigtext{CodeBERT} and \CodeT \layerperfbigtext{CodeT5} encode the probed task characteristics in the middle layers. All models are \BERT variants. Interestingly, \GraphCodeBERT has a very pronounced pattern, with very strong middle layers, while layer 10 is almost always one of the worst performing layers. 

\myparagraph{Late layers.} \CodeReviewer \layerperfbigtext{CodeReviewer} and \CodeBERTa \layerperfsmalltext{CodeBERTa} show a progressive learning pattern with the best performance generally originating in the last layer of the model. Similarly, \BERT \layerperfbigtext{BERT} show a progressive learning pattern, but since the average performance drops in the last layer, the penultimate layer encode the most information. \PLBART \layerperfsmalltext{PLBART} clearly performs best in the second half of the layers, but the best performing one is closer to the middle.

In contrast, \UnixCoder \layerperfbigtext{UnixCoder} shows a learning pattern which is unique, with several individual layers much better than their neighbours (especially layers 2, 8, and 12), with layer 8 performing best overall.

\vskip 2mm %
\noindent
It is a common misnomer that the final layer of the model is the best layer. However, in reality we observe a diversity learning patterns for source code models. Indeed some models progressively improve over the layers with the final layer producing the best results, yet, we also notice some models encoding significant information predominantly in the middle layers, as well as in the first layers.

\resultbox{\textit{Source code models show a variety of learning patterns, and do not always perform best in the last layer.}}

\section{Discussion}
\label{sec:discussion}

\subsection{Model performance} \RS{Overall, both} \CodeBERT and \GraphCodeBERT \RS{shows promising results: they are the most consistent models, improving upon the} \BERT \RS{baseline in almost all of the tasks. In fact, they are the only two models to consistently improve over} \BERT \RS{on the \emph{structural} tasks, as well as being the only ones to offer sizeable improvements over} \BERT \RS{in a challenging task such as} \JBL. \GraphCodeBERT underperforms the baseline only on \LEN. We hypothesize that since the \LEN task does not depend on source code structure, or on the data-flow, the modelling paradigm in \GraphCodeBERT extracts the surface-level information relevant to the number of tokens less effectively compared to other models. A deep dive into the task analysis promises to yield more concrete evidence. \CodeBERT \RS{underperforms on} \SRK; \RS{while we have no clear hypothesis of why that is, we note that this is the least source code specific of the three} \switchbased \RS{tasks.}

\subsection{The competitiveness of \BERT}
We chose to include \BERT as a baseline specifically for its \emph{lack} of source code knowledge. \LEN is only task not about source code per se; thus \LEN was the only task where we expected \BERT to be competitive. Also, our expectations could have been that \BERT might do well on tasks that rely on identifiers. Identifiers are predominantly composed of concatenated English words, so \BERT might have been able to use its English knowledge. We expected it to underperform on all code-specific tasks, particularly the structural ones, for which source code differs most from English. 

In fact, \BERT’s knowledge of English did not particularly help it for tasks \RS{relying on identifiers such as \texttt{IDN} or \texttt{SRI}. }
However, it showed competitive performance for structural tasks. But this surprising performance on structural tasks is not due to \BERT performing well; rather, it is more due to the source code models performing surprisingly \emph{poorly}.  

Clearly, all source code models struggle on structural tasks. One possible reason for this is that pre-trained source code models using the Transformer architecture do not fully exploit the structure of source code. Transformers inherit by default positional embeddings, popular for \texttt{NLP} tasks, that emphasize the sequential nature of tokens. While this can be changed, few source code models do it, or adopt a training objective that takes into account the specificities of source code (a point discussed further in Section~\ref{sec:factors}).

\subsection{Factors influencing performance}
\label{sec:factors}

We have included several models in this study, but each model varies in several ways from the other models, which prevents us from unambiguously identifying the factors that could influence the results. We nevertheless outline our hypotheses for several such factors.

\myparagraph{Influence of the language.} One hypothesis could be that models trained on a single language (\textit{Java}-specific) could perform better, since they do not have to dedicate capacity to other programming languages. We find no evidence of this, so far. In fact, monolingual models are among the worse performing models in this study (\JavaBERT, \PLBART\footnote{\textit{The \emph{\PLBART} model checkpoint we have probed is trained specifically on \textit{Java} on top of the \emph{\PLBART} base model which is multilingual.}}). But a real test of this would be, for instance, to compare a monolingual \GraphCodeBERT with a multilingual one.

\myparagraph{Influence of the training data.} Unlike the other models, \CodeReviewer is trained on diff changes rather than functions. It performs best on \KTX, \IDN, \LEN, and, to a lesser extent, on \REA tasks. One could think that these identifier level tasks are closer to its training data. On the other hand, it struggles with tasks that require to reason on an entire function (e.g., \CPX, \VCU), which are understandably farther from its training data.

\myparagraph{Performance of decoders.} We initially also wanted to include \textit{decoder-only} models, such as \texttt{CodeGPT} or \texttt{CodeGen}. This would have been especially valuable as the largest models such as \texttt{Codex} or \texttt{AlphaCode} are decoder only models (both are not available for this study: \texttt{AlphaCode} is not accessible, and Codex does not allow us to access model weights). However we found these models to be under-performing. One possible reason could be that decoders function differently than encoders, and thus could be harder to probe. \RS{The performance on }\LEN \RS{points (albeit weakly) to encoder/decoders behaving differently already: three encoder/decoder outperform the }\BERT \RS{baseline by a sizeable margin, while all the encoders are close to }\BERT\RS{ or underperform it. Of note, }\UnixCoder, \RS{which has encoder/decoder capabilities, but that we use in encoder-only mode, does not outperform} \BERT. A second reason could be that these models are trained on a classical language modelling objective as a continuous stream of \emph{files}, rather than encoding single \emph{functions}. As such, they may have a very distinct vision of their input, and not have such a clear boundary of functions, which are purposed as probe samples.

\myparagraph{Influence of Model size.} It is hard to extrapolate from the limited data. The best performing models (\CodeBERT, \GraphCodeBERT) are not particularly large. \CodeT is among the largest models, and performed relatively well. \CodeReviewer is also larger, but performed less well than \CodeT, although the different training modality played a role. Similarly, \PLBART is somewhat larger than \CodeBERT or \GraphCodeBERT, and yet is one of the worst performing models. On the other hand, the two other under-performing models, \JavaBERT and \CodeBERTa, are on the smaller end. Another aspect is the number of layers: both \CodeBERTa and \PLBART have only 6 layers, while the other models have 12. Overall, there is insufficient evidence at this stage to determine how model size would impact performance. A stricter approach with probes that fixes all parameters but varies model size would be necessary to study the influence of model size on encoding of intrinsic code characteristics.

\myparagraph{Influence of the training objective.} We note that all the models that under-perform on the \BERT baseline have a training objective that is derived from \texttt{NLP}, but is not source code specific. On the other hand, \GraphCodeBERT’s training objective forces it to reason about the data flow; \CodeT has several identifier-specific pre-training tasks; \UnixCoder uses \texttt{AST} information while training; while \CodeReviewer has training objectives related to code changes. The only exception is \CodeBERT. \CodeBERT \RS{uses a pre-training objective (RTD---Replaced Token Detection) that is not \emph{explicitly designed} for source code. On the other hand, RTD can \emph{implicitly} provide a very source-code specific objective. With RTD,} \CodeBERT \RS{learns to discriminate between real source code and fake, but plausible, source code (source code where some tokens are replaced by the output of a bidirectional n-gram model} \cite{feng2020codebert}). \RS{We also note that} \GraphCodeBERT \RS{performs comparatively well on the} \VCU \RS{task, which is close to its training objective (reasoning on the data flow of variables). The fact that models that have language-specific objectives (explicitly or implicitly) tend to overperform, while more generic models underperform, constitutes in our view evidence that the training objective is a key factor in the model's performance, although further studies should confirm this. We note that Troshin and Chirkova's study (discussed in Section}~\ref{sec:otherstudies}) \RS{points towards similar conclusions.}

\subsection{Additional Observations}

\myparagraph{Performance on structural tasks.} We also examine the confusion matrices to better understand the performance of models on the tasks. For the \countbased or \compbased tasks, we note that the models perform much better at finding the class ``0'' than any other classes. It appears that models can differentiate well between the absence of a phenomenon and its presence, but counting is farther out of reach. The models also tend to over-represent one class with a low count, and another one with a higher count, showing some ability to differentiate between low and high counts. 

\myparagraph{From 1K to 10K.} We also examined the change in performance when training on only 1,000 samples, rather than 10,000. On average, the models improve their accuracy by close to 5\% when training on 10,000 samples. While this is good, this also means that the models 
need a significant number of samples (at least 10K) for effective learning, which means that the representations are not that easy to access. The best performing models improved the most (\GraphCodeBERT, followed by \UnixCoder and by \CodeBERT). Another crucial aspect observed when training the probes with 1,000 and 10,000 samples is that the irregularity (uneven trend) in accuracies across the model layers is stabilized as more samples are provided. This gives a more robust characterization of the model learning for the different intrinsic code characteristics. We have observed this for all tasks in our probing suite.

\subsection{Implications}

\myparagraph{Using \texttt{NLP}-based models as baseline.} We were surprised by \BERT’s performance in some tasks, which provides very valuable context for the performance of the source code models. As such, comparing with a \texttt{NLP}-based baseline that is not specifically trained on code, and that is \emph{not} expected to perform well on source code, may yield unexpected results, and help researchers better frame their results.

\myparagraph{\INSPECT extension.} We defined a suite of probing tasks, but this is by no means final. Our framework allows for addition of new tasks and new languages. Then they can be used to run additional experiments. For instance, releasing similar tasks in other programming languages would allow to test whether multi-lingual model have similar quality of representations in all the languages that they were trained on. \RBtext{R2.C7.1}{Extending inspect to another language would not drastically change inspect, but significant data gathering and pre-processing might need to take place before this.} \RBback Additional tasks would allow to probe for representations of additional concepts that our initial suite could not cover, and potentially shine a light on additional shortcomings of source code models. We have in mind several possible tasks that could be probed but did not include as of now (see \RA{Appendix} ~\ref{appendix:tasks:candidate}), and we have no doubts the community would have many more. \RAtext{R1.C2.2}{Another way to extend our framework is to support model architectures beyond Transformers. This would allow us to probe for additional models that represent more explicitly the structure of programs in their architecture, such as GNNs} \cite{DBLP:journals/corr/abs-1711-00740}, \RA{models that use path-based representations such as Code2Seq}\cite{alon2019code2seq}, \RA{or models that augment path-based representations with precise flow information such as Flow2Vec} \cite{sui2020flow2vec} \RA{and ContraFlow} \cite{cheng2022path} \RAback. 

\myparagraph{Further research on training objectives.} Our results provide evidence that the training objective influences the model’s representation of source code concepts. Moreover, our results show that if models perform well on semantic tasks that can be solved by reasoning at the level of tokens or sequence of tokens, the models struggle with probing tasks that require them to reason about source code structure. This holds even for the best performing models such as \GraphCodeBERT. Clearly, this calls for additional investigation into whether training objectives that take the structure of code into account would impact performance, and in which way. For instance, it is possible that such training objective would come with tradeoffs: if \GraphCodeBERT is one of the best models in the structural tasks, it is one of the worst models on \KTX, for instance. Designing novel training objectives, investigating if such tradeoffs occur, and if they can be mitigated, are all ample avenues for future work. \RAtext{R1.C2.1}{Some particularly interesting training objectives to investigate include ones that model program paths, such as Flow2Vec}\cite{sui2020flow2vec}, \RA{that leverage contrastive learning such as ContraFlow} \cite{cheng2022path}, \RA{or approaches that combine multiple models such as Fix-Filter-Fix} \cite{hong2021fix}. \RA{However some of these approaches would need INSPECT to be extended}\RAback.

\myparagraph{Transfer to downstream tasks.} Finally, our probing task suite investigates only the pre-training stages of source code models. Whether better representations after pre-training translate to increased performance on end tasks, or if fine-tuning is enough should be systematically investigated. 

In addition, our task suite could be used to investigate to which extent fine-tuned models conserve their representations of the probed concepts, or if they are subject to ``catastrophic forgetting''. \RB{Troshin and Chirkova have some initial results that point towards this in their study}\cite{troshin2022probing}. Finally, we note that the paradigm for the largest language models such as Codex or AlphaCode is at the moment to forego or de-emphasize fine-tuning and to rely on prompt engineering instead. While we do not have access to the inner weights of these models to probe them, such a use case constitute an additional incentive to probe these models, and to perform research in finding the right pre-training objective for source code models.

\myparagraph{\RCtext{R3.C1}{Towards a better understanding of source code models.}} \RC{Diagnostic tasks, taken in isolation or together as a whole, can be used to further our scientific understanding of what source models learn. In a context where these models are both notoriously opaque, but, at the same time, taking an increasingly larger place in the world, we think that this is an important endeavor. A very interesting direction for future work is to study the link between diagnostic tasks and practical tasks. Practical tasks are more complex than diagnostic tasks, which probe for very specific characteristics. However, we hope that diagnostic tasks can be used to form hypotheses on the performance of practical tasks (e.g., that performance on a practical task might be subpar on some parts of the data due to a model's lack of structural understanding). Needless to stay, extensive further work in this direction would be needed to confirm that it is practical.} \RCback

\subsection{Relationship to other studies}
\label{sec:otherstudies}

 As mentioned in Section~\ref{sec:background}, several studies have also analyzed source code models since our first study. We briefly discuss their results in relation with ours.

\myparagraph{\RB{Comparison with Troshin and Chirkova's study.}} \RBtext{R2.C6.2}{We find both some similarities and some differences, which we discuss here. At a high level, we find that the models performed considerably worse on our tasks (e.g. in \texttt{Algorithm} the ``worse’’ task from Troshin and Chirkova, some source code models exceed 75\% accuracy). This could be due both to the selection of tasks (we are testing for source code characteristics that are less well represented), or the data selection (starting with more data, we were able to select a more challenging subset, e.g. enforcing that it is balanced).}

\RB{Since our selection of tasks is different, we can not do a detailed per-task comparison. However, we note that there is a degree of similarity between their \texttt{Variable Misuse} task, and our \texttt{SRI} task. In both tasks, there is a considerable gap between \texttt{BERT}’s performance and the best-performing models. In both tasks, \GraphCodeBERT, \CodeBERT, and \texttt{CodeT5} perform well (although \texttt{CodeT5} performed less well on \texttt{SRI} task, and \texttt{PLBART} performed considerably worse—it seems that our version of the task is more discriminative). }

\RB{Turning to the high-level conclusions, Troshin and Chirkova observe that ``[\texttt{BERT}] performs worse than the models pretrained on code in all tasks except the semantic-related \texttt{Readability} and \texttt{Algorithm} tasks, where all pretrained models perform similarly’’ (one caveat is the size of the datasets for \texttt{Algorithm} and \texttt{Readability}, 934 and 200 datapoints). We find that we have significantly more tasks where \texttt{BERT} performs comparably or better than some or most pre-trained models; in fact, \texttt{BERT} achieves third place in 4 occasions out of 15 tasks, including all tasks related to code complexity. In only 7 tasks out of 15, we find that the majority of source code models do better than \texttt{BERT}. If both studies point out strengths and weaknesses of source code models, our study has more conservative conclusions.}

\RB{Troshin and Chirkova also consider whether models pretrained with code-specific objectives perform better. For instance, they find that \texttt{GraphCodeBERT} performs best in their \texttt{DFG Edge Prediction} task, which is similar to its pretraining objective. Likewise, we find that \texttt{GraphCodeBERT} performs best in our \texttt{VCU} task, which bears some similarity to its pretraining  objective. On difference is that we observe consistently worse performance for models with basic NLP objectives (some of them not in Troshin and Chirkova’s study), which provides us with additional evidence supporting the conclusion, but from the other side of the spectrum. Overall, the fact that both studies find evidence that code-specific pre-training objectives have a positive impact on probing performance increases our confidence in this conclusion}.

\RB{Troshin and Chirkova also investigate performance across layers (discussed below). Finally, they also consider model size: they were able to compare two variants of the \texttt{CodeT5} and \texttt{PLBART} models on their task suite. However, they find mixed evidence of the benefit of model size (it is positive in 6 of 8 tasks for \texttt{CodeT5}, and 4 of 8 tasks for \texttt{PLBART}). This echoes our lack of conclusion on model size. } \RBback

\myparagraph{Layer-wise performance}. Several studies have found that structural information was best represented in the middle layers. This is the case of the \texttt{AST}-Probe of Lopez \etal, which finds that the representation of the \texttt{AST} is better defined in the middle layers \cite{lopez2022ast}. This find is echoed by the study of Wan \etal \cite{wan2022they}, who also measure the representation of the \texttt{AST} by source code models. Our findings for the structural tasks go in this overall direction, as we find that models tend to perform best with the middle layers for these tasks. However we also encountered several tasks for which early or late layers where the ones where the models were performing best, indicating that not all tasks are behave similarly.  Troshin and Chirkova also investigate performance across layers; they find that for most tasks, the middle layers (4 to 10) have the highest performance \cite{troshin2022probing}. They also find that some tasks related to variable names and misuse perform best on the last layer. We also find that structural tasks perform best in middle layers, and that semantic tasks tend to perform best in the last layers. The work by Chen \etal \cite{chen2022cat} defines the \texttt{CAT} metric, which performs best in the early layers; however, the \texttt{CAT} score is based on the attention. 

\myparagraph{Model performance.} Looking at the performance of specific models, the study of Lopez \cite{lopez2022ast} compares several models in common with ours. We note that our results go in the same direction: For both \CodeBERT and \GraphCodeBERT, the AST-Probe metric peaks in earlier layers than \CodeT. The shape of the AST-probe curve for \CodeBERTa is also similar to our results. We also note that \CodeBERT and \GraphCodeBERT perform the best with respect to the AST-Probe, which, again, echoes our results. In Wan's study \cite{wan2022they}, \GraphCodeBERT outperforms \CodeBERT. In terms of CAT-Score \cite{chen2022cat}, \GraphCodeBERT is usually ahead, followed by \CodeBERT, while \UnixCoder struggles to compete with a RoBERTa baseline. In Troshin and Chirkova's probing tasks, \GraphCodeBERT is often, but not always the best performing; \CodeT is often competitive and performs the best in their ``is variable declared'' and ``algorithm'' task; \CodeBERT is often competitive as well. They also find that \BERT is competitive in the ``readability'' and ``algorithm'' tasks. Finally, if somewhat less related, another work observed surprisingly good performance for \BERT on source code tasks, when \BERT is extended with source-code specific adapters \cite{goel2022cross}.

Overall, our findings contribute to a growing body of evidence about the way source code models represent various source code aspects, and in which layers they do so. Our findings are in general in agreement with other studies. However, our extensive selection of tasks and our comparison with the \BERT baseline allows us to highlight the limitations of the current crop of source code language models, as well as suggesting ways forward in terms of new experiments, and potential solutions with new training objectives.


\section{Limitations}
\label{sec:limitations}
In this section, we discuss some of the limitations of our study. These may be addressed in future works as we collectively develop the probing paradigm further, ultimately, to better understand the inner workings of large language models (\texttt{LLM}s), particularly in our case, large-scale source code language models.

\myparagraph{Mono-lingual probes.} The majority of the probed models are pre-trained on multiple programming languages, and probing them on a single language (e.g. \textit{Java}) is just a first step. Further probes in more languages should be designed to understand the code aspects learned by the pre-trained models. Our work presents the probing framework and releases the relevant code with which different models can be probed on other languages. \RBtext{R2.C7.2}{The work to gather and pre-process the necessary data in other languages should still be done, however, and is likely to require some effort.}\RBback

\myparagraph{Possible confounding factors.} While we did our best to make sure our tasks are reliable, there can be an influence from confounding factors. In particular, we can think of method size as a possible confounding factors for some of the structural tasks, as it is well known that there is often a relationship between lines of code and source code complexity in general \cite{graylin2009cyclomatic}. This is why we chose to formulate the task as a classification task, rather than a regression task. We think that the impact of this compounding factor, if present, is limited, since we observe very different performance for \LEN compared to the \compbased and \countbased tasks (e.g., \PLBART is one of the best performing models for \LEN, but one of the weakest on the \compbased and \countbased tasks). Moreover, should this confounding factor be more significant, this would only further highlight that the source code models, by overly relying on length, have even weaker abilities to model source code structure than we previously thought.

\myparagraph{Unknown characteristics.} We limited this study to fifteen tasks, in an attempt to balance the broadness of characteristics necessary to study source models, and the need to reduce the complexity of the presentation. Nevertheless, code understanding may depend on several source code characteristics not covered in this study; further exploration beyond probing for our current task suite is needed to explore additional characteristics. We present a few possibilities for further probing tasks in Appendix ~\ref{appendix:tasks:candidate}, as well as some tasks that we excluded to streamline the presentation.

\myparagraph{Method-level code.} Our tasks all focus on the method as a unit (save for \IDN which focuses on single identifier tokens).  We note that broader contexts beyond methods are still challenging for all but the largest source code models \cite{DBLP:journals/corr/abs-2106-15209-small-data} \cite{karmakar2022jemma}, due to the limited input size of the transformer. This is why we restrict ourselves to methods only. Tasks that encompass a larger context should, in time, be developed.

\myparagraph{Focus on source code.} Our tasks primarily focus on source code, while many models can handle both source code and natural language. We chose to focus primarily on source code, since probing natural language itself has been extensively investigated \cite{rogers2020primer, belinkov2022probing}. However, additional probing tasks that investigate both source code and natural language would be a worthwhile addition.

\myparagraph{Focus on encoders rather than decoders.} We probed two decoder-only models, CodeGPT \cite{lu2021codexglue} and CodeGen  \cite{nijkamp2022conversational}. However, our initial results found that these models were severely under-performing, so we preferred to exclude them from the study as we suspected more factors could be at play (see Section~\ref{sec:discussion}). While the current study is specifically probing for \textit{encoded} code characteristics in the model, probing on vector representations from decoders in the future may yield interesting results. Even though early evidence indicates that vectors from the encoders extract more information than the decoders \cite{troshin2022probing}, yet, probes on decoders may be useful, and perhaps more fair for certain types of models.

\myparagraph{Model availability.} Our study is subject to the availability of suitable models to probe. We can not control which models are trained, nor how they were trained. This necessarily limits our conclusions, as the models that are available may not have all the characteristics that we would like to probe for. For instance, we would have liked to probe more mono-lingual models, but the research community has principally focused on multi-lingual language models. We would have also liked to perform more systematic study of model sizes, but this is also limited by model availability. The fact that each model varies in multiple factors prevents us, at this time, to emit strong conclusions as to which factors influence performance. Likewise, the largest and most successful source code language models (e.g., Codex, AlphaCode) are not available with the necessary degree of access to include them in our study.

\section{Conclusion}
\label{sec:conclusion}
Large-scale pre-trained models for code have been shown to perform spectacularly well on a range of Softwae Engineering (SE) tasks leading to the release of a number of popular new tools e.g. Tabnine, IntelliCode \cite{svyatkovskiy2020intellicode}, or Github Copilot \cite{chen2021evaluating}, among many others. As more of such pre-trained models and derivative tools are introduced to the SE community, it becomes imperative to improve our understanding of their capabilities and weaknesses. 

In this paper, we have used the probing paradigm to gain insight into the capabilities of eight state-of-the-art publicly-available pre-trained source code models. We gauged their capabilities on a set of fifteen tasks specifically designed for this study to evaluate a broad set of source code characteristics, including identifiers, structural, and semantic characteristics. We show how probes can help us uncover the strengths and weaknesses of a model, to understand the role played by the individual hidden layers in model performance, to verify the linear extractability of properties, and overall to peek into the ``black boxes'' that are large-scale pre-trained models.

In summary, we observe that \GraphCodeBERT is best performing model across most tasks, encoding more of the syntactic, semantic, and structural information than any other model. While it is hard to isolate all factors unambiguously, we think that the training objective is one of the most important factors that impact performance.

More importantly, we notice that models struggle with structural tasks. Even \GraphCodeBERT's improvement over \BERT---a model that should \emph{not} have much source code knowledge---is slim. This suggests that there is room for further research in architecting more advanced source code models that can more effectively leverage source code knowledge. Additionally, we observed a diversity of learning patterns in the model layers, indicating that care must be taken as to determine which layers encode the most knowledge for specific tasks. 

We introduced a probing framework, \INSPECT, for the intrinsic evaluation of large-scale pre-trained models of source code. Our probing framework automates the entire probing process and can be used with any model available on the Huggingface model hub. Furthermore, it is also extensible with additional tasks. As future work, we plan to construct probing datasets in multiple languages since the majority of the pre-trained code models are multi-lingual. We also plan to define additional tasks to cover more source code characteristics.

In the long run, such an extensive suite of probing tasks could be used to thoroughly evaluate novel pre-trained source code models, thereby forming a pseudo-benchmark during the development phase, making sure that these models do encode important source code characteristics.

\bibliographystyle{IEEEtran}
\bibliography{references}

%

\newpage

\begin{IEEEbiography}
[{\includegraphics[width=1in,height=1.25in,clip,keepaspectratio]{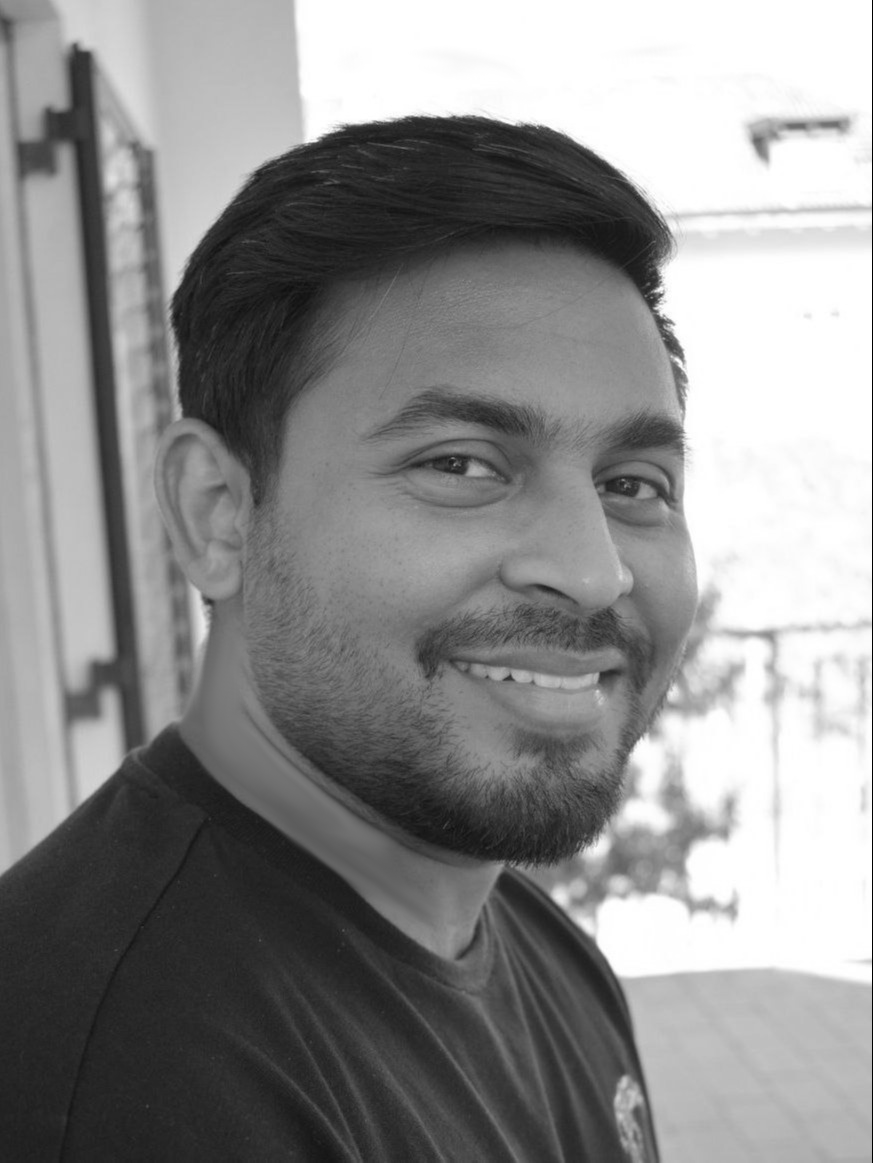}}]{Anjan Karmakar} received his Ph.D. and M.Sc. in
Computer Science from the Free University of Bozen-Bolzano, Italy. His research interests include applications of machine learning in Software Engineering, and evaluation of Large Language Models (LLMs) of source code. 
\end{IEEEbiography}


\begin{IEEEbiography}
[{\includegraphics[width=1in,height=1.25in,clip,keepaspectratio]{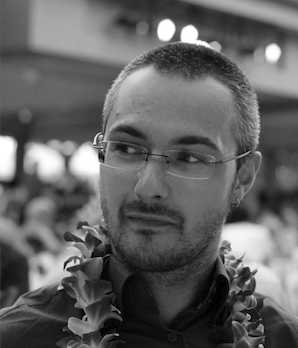}}]{Romain Robbes} is a Senior Scientist (Directeur de Recherche) at the CNRS, working at the LaBRI and hosted by the University of Bordeaux, since February 2023. Before that, he was: an Associate Professor at the Free University of Bozen-Bolzano, (2017–2023); an Assistant, then Associate Professor at the University of Chile (2010–2017); a Ph.D. student, then post-doc, at the University of Lugano (2004–2009). His research interests include Empirical Software Engineering, Mining Software Repositories, Software Maintenance and Evolution, and the intersection of Machine Learning with Programming Languages and Software Engineering.
\end{IEEEbiography}

\vspace{\fill}

\pagebreak







%


\newpage
\appendices
\section{}
\label{appendix:heatmaps}

\noindent\begin{minipage}{\textwidth}
    \centering
    \captionof{figure}{Heatmaps illustrating layer-wise model performance across tasks}
    \includegraphics[width=0.95\textwidth]{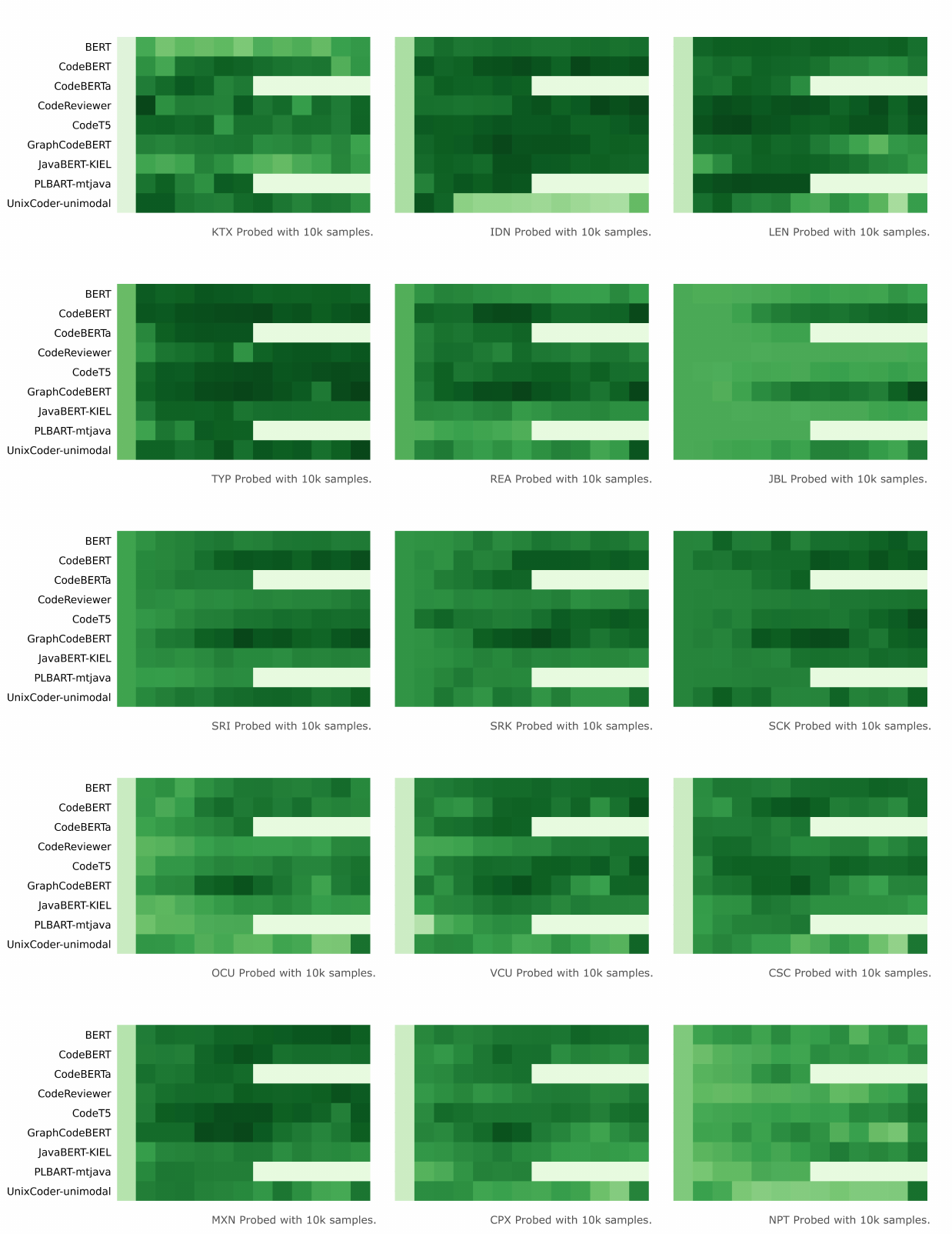}
\end{minipage}

\clearpage

\section{Candidate tasks}
\label{appendix:tasks:candidate} 
We briefly describe tasks that we considered for inclusion, but finally did not include, as well as tasks that we could consider including in the future.

\subsection*{Discarded tasks}
During the course of this work, we developed and experimented with several tasks, but the following are some of which did not make the final crop of probes.

\myparagraph{\AST and \TAN.} Our previous work \cite{karmakar2021probes} included a single token-level \texttt{AST} Node Tagging task. It mixed identifiers with keywords, and did not test for keyword generalization. We replaced it with the current token-level tasks: \IDN and \KTX. This allows to separately probe keyword and identifier tagging, which are more nuanced than \texttt{AST} node tagging. 

\noindent
As for \TAN, it was a binary classification task where models were made to predict the most common \texttt{AST} node type (between identifiers and keywords) in a method. However we found that all models performed well on the task, which led us to de-prioritize it in order to simplify the presentation.

\myparagraph{\JMB and \JFT.} We experimented with two variants of \JBL. While \JBL swaps a single token pair, the variants swapped 50\% (\JMB) and 100\% (\JFT) of the tokens. Both variants were too easily solved by the models, indicating that the models are able to detect such drastic code changes.

\myparagraph{\OCT and \VCT.} They were variants of \OCU{} and \VCU that focused on the total count of operators and variables, rather than the unique count. We chose the unique variant of each as we thought it was less sensitive to the confounding factor of code size, and to reduce the number of tasks; .

\myparagraph{\NML and \NMS} They are descendent tasks of \CSC where the number of loops (\NML) and the number of if structures (\NMS) are probed separately in each task. The predictions from the \NML and \NMS tasks were altogether aligned with the results from the \CSC task, therefore, we decided to only include \CSC to minimize the number of similar tasks.

\subsection*{Possible Future tasks}
We also discussed additional tasks, that we have not included at this time to limit the amount of tasks. 

\myparagraph{\texttt{AST} level tasks.} We thought that characteristics based on the source code's \texttt{AST}, such as \texttt{AST} depth, could be a valuable addition. For the moment, the code complexity tasks cover the most similar concepts. Similarly, we considered classifying methods according to some patterns in the AST structure (e.g., ``has nested loops'', ``has an if in a loop'', \emph{etc.}). However, we thought the diversity of possible patterns might be too high, so we used complexity metrics instead.

\myparagraph{Variant incorrect-code tasks.} We considered a variant of \SRI were \emph{all} the occurrences of an identifier would be swapped with a random identifier, further emphasizing sensitivity to context. However we thought such a task might be better suited as a future work. We also considered higher level variants of the ``code jumbling'' tasks, such as jumbling statements or entire code blocks. These would be extremely interesting, but we thought that they might be too complex for the current breed of models.





\ifCLASSOPTIONcaptionsoff
  \newpage
\fi



%



\end{document}